\documentclass[11pt, letterpaper]{article}

\usepackage{fullpage}
\usepackage{amsmath}
\usepackage{amssymb}
\usepackage{hyperref}
\usepackage{xcolor}
\usepackage{parskip}
\usepackage{graphicx}
\newcommand{\nn}{\nonumber}
\newcommand{\ba}{\begin{align}}
\newcommand{\ea}{\end{align}}
\newcommand{\be}{\begin{equation}}
\newcommand{\ee}{\end{equation}}
\newcommand{\bea}{\begin{eqnarray}}
\newcommand{\eea}{\end{eqnarray}}
\newcommand{\bp}{\begin{pmatrix}}
\newcommand{\ep}{\end{pmatrix}}
\newcommand{\bi}{\begin{itemize}}
\newcommand{\ei}{\end{itemize}}
\newcommand{\overbar}[1]{\mkern 1.5mu\overline{\mkern-1.5mu#1\mkern-1.5mu}\mkern 1.5mu}
\newcommand{\m}{\mathcal}

\newcommand{\D}{\mathcal{D}}
\newcommand{\oD}{\overbar{\mathcal{D}}}
\newcommand{\p}{\partial}

\def\pa{\partial}

\def\ep{\varepsilon}

\def\m{\mu}
\def\n{\nu}

\def\p{\phi}

\def\th{\theta}
\def\tb{\bar{\theta}}

\def\pt{\partial}

\def\Db{\bar{D}}

\def\P{\Phi}
\def\Pd{\Phi^\dag}

\newcommand{\eps}{\epsilon}
\newcommand{\M}{M_{P}^2}
\newcommand{\K}{\tfrac{K}{\M}}

\begin{document}

\bigskip
\thispagestyle{empty}
\bigskip
\begin{center}
{\LARGE \bf Supergravitational Conformal Galileons}
\end{center}

\bigskip
\bigskip
\begin{center}
{\large {\bf Rehan Deen} and {\bf Burt Ovrut}}
\end{center}

\begin{center}
    {\it Department of Physics and Astronomy\\
     University of Pennsylvania \\
     Philadelphia, PA 19104--6396}\\
\end{center}

\bigskip
\bigskip
\bigskip
\begin{center}
\textbf{Abstract}\end{center}
{\noindent
}
The worldvolume actions of 3+1 dimensional bosonic branes embedded in a five-dimensional bulk space can lead to important effective field theories, such as the DBI conformal Galileons, and may, when the Null Energy Condition is violated,  play an essential role in cosmological theories of the early universe. These include Galileon Genesis and ``bouncing'' cosmology, where a pre-Big Bang contracting phase bounces smoothly to the presently observed expanding universe. Perhaps the most natural arena for such branes to arise is within the context of superstring and $M$-theory vacua. Here, not only are branes required for the consistency of the theory, but, in many cases, the exact spectrum of particle physics occurs at low energy. However, such theories have the additional constraint that they must be $N=1$ supersymmetric. This motivates us to  compute the worldvolume actions of $N=1$ supersymmetric three-branes, first in flat superspace and then to generalize them to $N=1$ supergravitation. In this paper, for simplicity, we begin the process, not within the context of a superstring vacuum but, rather, for the conformal Galileons arising on a co-dimension one brane embedded in a maximally symmetric $AdS_{5}$ bulk space. We proceed to $N=1$ supersymmetrize the associated worldvolume theory and then generalize the results to $N=1$ supergravity, opening the door to possible new cosmological scenarios.
\vskip 7cm
\hrulefill

Email:~~rdeen@sas.upenn.edu,~ ovrut@elcapitan.hep.upenn.edu

\newpage 

\section{Introduction}

This paper is intended as a preliminary step to accomplish the following; first, to present a method for extending the bosonic worldvolume theories of 3+1 dimensional probe branes embedded in non-dynamical bulk spaces to flat $N=1$ supersymmetry--both in superfields and in the associated component fields, and second, once this has been accomplished, to couple such worldvolume theories to $N=1$ supergravity, thus allowing for curved spacetime as well as gravitational dynamics. Here, we will carry this out within the relatively straightforward context of a three-brane embedded in a maximally symmetric $AdS_{5}$ bulk space. This 3+1 brane bosonic worldvolume theory is, as we will discuss below, already known to produce the theory of conformal Galileons. Hence, in this work we will be explicitly computing the $N=1$ supersymmetric extension of conformal Galileons in flat superspace and then generalizing them to $N=1$ supergravity.

The worldvolume action, and the associated dynamics, of a bosonic 3+1 brane embedded in a background five-dimensional bulk space are of considerable interest \cite{Dvali:2000hr}. To begin with, the structure of the worldvolume theory itself has been shown to possess remarkable topological and dynamical properties, depending on the symmetries of the bulk space. For example, it was demonstrated in \cite{deRham:2010eu} that a probe three-brane embedded in an maximally symmetric $AdS_{5}$ space led to the theory of relativistic DBI conformal Galileons which, in the low momentum limit, reproduced the five conformal Galileons first discussed in \cite{Nicolis:2008in}. This was generalized in \cite{Goon:2011qf, Goon:2011uw} to different maximally symmetric bulk spaces--including, $M_{5}$ and $dS_{5}$. These bosonic brane worldvolume actions were shown to contain new effective field theory generalizations of Galileons, each reflecting the background symmetry groups imposed on them.
In addition to the novel 3+1 dimensional effective field theories discovered in this manner, bosonic three-branes embedded in a higher dimensional bulk space can lead to new and exotic theories of cosmology and the early universe. For example, it was demonstrated in \cite{Creminelli:2010ba, Hinterbichler:2012yn} that the worldvolume theory of a three-brane moving relativistically in an $AdS_{5}$ background--that is, the DBI conformal Galileons-- can, for an appropriate choice of coupling parameters, admit a stable, Poincare invariant background that violates the Null Energy Condition (NEC). This allows for a cosmological theory in which the Universe begins as a non-singular flat geometry and then ``expands'' to the Universe that we observe--so-called Galileon Genesis. The fact that bosonic brane worldvolume theories can, under the appropriate circumstances, admit NEC violation, has also led to ``bouncing'' cosmological scenarios \cite{Creminelli:2007aq, Cai:2012va, Brandenberger:2012zb, Brandenberger:2016vhg, Koehn:2013upa, Koehn:2015vvy, Easson:2011zy, Ijjas:2016tpn, Ijjas:2016vtq}. In these, a contracting Friedman-Robinson-Walker (FRW) geometry can bounce smoothly through the ``Big Bang'' to the present expanding spacetime.

Although these bosonic braneworld scenarios are interesting, the fact remains that branes of varying dimensions embedded in higher-dimensional bulk spaces arise most naturally within the context of supersymmetric string theory and $M$-theory. Furthermore, whereas the spectrum and interactions of particle physics must simply be added in an ad hoc manner to bosonic cosmological scenarios, it is well-known that the Standard Model can arise  as the spectrum of specific superstring vacua that simultaneously include various types of branes. One very concrete example is the compactification of $M$-theory to five-dimensions known as Heterotic M-Theory \cite{Lukas:1998yy}. In this theory, the particles and interactions of the Standard Model arise on the so-called ``observable'' wall \cite{Lukas:1997fg, Braun:2005nv, Buchbinder:2002pr} of an $S_{1}/{\mathbb{Z}}_{2}$ orbifold, whereas a ``hidden sector'' composed of unobserved particles occurs on a second orbifold wall--separated from the first by a specific five-dimensional geometry \cite{ Lukas:1998tt}. Naturally embedded within this five-dimensional bulk space are 3+1 branes (five-branes wrapped on a holomorphic curve), whose existence is required for anomaly cancellation and, hence, consistency \cite{Lukas:1998uy}. In addition to this natural setting for particle physics and 3+1 brane worldvolume theories, there is a second, very significant, new ingredient. That is, these vacua, prior to possible spontaneous symmetry breaking, are all $N=1$ supersymmetric. 
These realistic vacua of supersymmetric three-branes embedded in heterotic $M$-theory led to the postulation of the ``Ekpyrotic'' theory of early universe cosmology \cite{Khoury:2001wf}. In this theory, a relativistic three-brane embedded in the five-dimensional bulk space is attracted toward the observable wall via a potential energy, which arises from the exchange of $M$-theory membranes. This potential was explicitly computed in 
\cite{Lima:2001jc} and found to be a steep, negative exponential
\footnote{Within this context, the lowest order kinetic energy for the 3+1 brane position modulus was presented in \cite{Derendinger:2000gy}.}. Hence, in this phase, the universe is contracting. The scalar fluctuations of the brane modulus evolving down this potential produce two-point quantum fluctuations that are nearly scale invariant. As discussed in \cite{Khoury:2001bz}, under certain conditions the NEC can be violated and the universe ``bounces'' to the expanding spacetime that we presently observe. Furthermore, it was shown in \cite{Battarra:2014tga} that these fluctuations can pass through the ``bounce'' with almost no distortion and, hence, are consistent with observational data from the CMB. An effective field theory for the 3+1 brane modulus in the exponential potential was constructed in \cite{Buchbinder:2007ad}. However, the complete $N=1$ supersymmetric worldvolume action of the three-brane has never been explicitly constructed.
A first attempt to do this was carried out within the context of heterotic string theory in \cite{Antunes:2002hn, Ovrut:2012wn}. However, based on previous non-supersymmetric work \cite{Carter:1994ag, Bonjour:2000ca}, this was done by ``modelling'' the three-brane as a solitonic kink of a chiral superfield in the five-dimensional bulk space. Although some of the geometric terms, and particularly a computation of their coefficients, were found by these methods, the general theory of an $N=1$ supersymmetric three-brane worldvolume theory was far from complete. Given its potential importance in cosmological theories of the early universe, it would seem prudent to try to create a formalism for computing supersymmetric worldvolume brane actions in complete generality. In this paper, we begin the process of calculating these actions in a systematic fashion, starting with the bosonic actions discussed above, then supersymmetrizing them in flat superspace and then, finally, coupling them to gravitation by generalizing the worldvolume actions to $N=1$ supergravity. Specifically, we will do the following.

In Section 2, we review the formalism presented in \cite{Goon:2011qf, Goon:2011uw} for computing the bosonic worldvolume actions of 3+1 branes embedded in maximally symmetric bulk space geometries. 
First, the generic form of the five-dimensional metric is introduced in a specific coordinate system. We then present the general form of the worldvolume action composed of terms with two special properties; 1) they are constructed from worldvolume geometric tensors only and 2) they lead to second order equations of motion \cite{Deffayet:2009wt}. This restricts the number of such Lagrangians to five. Using the specific metric, we give the general form for four out of the five such Lagrangians--the fifth Lagrangian, ${\cal{L}}_{5}$, being very complicated and unnecessary for the purposes of this paper. In Section 3, again following \cite{Goon:2011qf, Goon:2011uw}, we review the four conformal DBI Lagrangians specifically associated with embedding the three-brane in a maximally symmetric $AdS_{5}$ bulk space. These Lagrangians are then expanded in a derivative expansion and all terms with the same number of derivatives assembled into their own sub-Lagrangians. Remarkably, as pointed out in \cite{deRham:2010eu, Goon:2011qf, Goon:2011uw}, these turn out to be the first four conformal Galileons. 

Section 4 is devoted to extending these four conformal Galileons from bosonic theories of a real scalar field $\phi$ to flat space $N=1$ supersymmetry. This was previously discussed in \cite{Khoury:2011da}, where the superfield Lagrangians for four of the five conformal Galileons were presented (the first conformal Galileon ${\cal{L}}_{1}$ was omitted). These four super-Lagrangians were then expanded into their component fields, two real scalars $\phi$ and $\chi$, a Weyl fermion $\psi$ and a complex auxiliary field $F$. However, this expansion was incomplete. In order to study the behaviour of the original real scalar field $\phi$, these super-Lagrangians were expanded to all orders in $\phi$ but only to quadratic order in all other component fields. There were two reasons for this. The first was to allow a discussion of some of the dynamics of the fermion field. The second reason was to permit a simple analysis of the complex auxiliary field $F$, which, to this order of expansion, does not contain higher-order terms in $F$ such as $(F^{*}F)^{2}$. These terms, along with the usual quadratic $F$ term,  were previously discussed in a non-Galileon context in \cite{Koehn:2012ar}. The associated cubic $F$ equation of motion was solved and a discussion given of the three different ``branches'' of the Lagrangian that now emerged. This work also looked into possible violation of the NEC and other ``bouncing'' properties in this context. Some interesting physics arising from these new branches was also discussed in \cite{Ciupke:2015msa}. However, in the present paper, we do something very different. Using the same superfield Lagrangian presented in \cite{Khoury:2011da}, we again expand into component fields--this time ignoring the fermion entirely, but working to all orders in the scalar fields. This opens up three very important new issues that will be presented and solved in this paper. The first arises due to the fact that the ${\cal{L}}_{1}$ bosonic Galileon had been ignored in the analysis of \cite{Khoury:2011da}. In this paper, we supersymmetrize ${\cal{L}}_{1}$, both in superfields and in component fields, and show that it leads to a specific potential energy in the theory. The second issue has to do with the stability of the two real scalar fields $\phi$ and $\chi$. This has two parts. First, one has to show that the potential energy so-derived, allows for stable solutions of the $\chi$ equation of motion. Related to this, one must show under what conditions the associated kinetic energy terms are non ghost-like. Both of these issues are discussed and solved in Section 4. The final issue that arises when one expands to all orders in the scalar component fields is, perhaps, the most important. It turns out that supersymmetric ${\cal{L}}_{3}$, when expanded to all orders in the component scalar fields, contains terms proportional to derivatives of the ``auxiliary'' field $F$--such as $\partial_{\mu}F$ and $\partial F^{*}\partial F$. Hence, it is no longer clear whether $F$ should be treated as an auxiliary field or as a dynamical degree of freedom. In this paper, we carefully discuss this issue and, within the context of a derivative expansion and a specific solution for $\chi$, solve for the $F$ field to leading, first and, finally, second order. To simplify the analysis, only the leading order results are inserted back into the full Lagrangian and the associated physics discussed.

Having carefully discussed the flat space $N=1$ supersymmetric conformal Galileons, we then extend the first three super Lagrangians, that is, supersymmetric ${\cal{L}}_{i}$,~$i=1,2,3$, to $N=1$ supergravity in Section 5. To do this, we expand upon the formalism previously discussed in \cite{Koehn:2012ar, Baumann:2011nm} as well as, within the context of new minimal supergravity, \cite{Farakos:2012je, Farakos:2012qu}. This is analytically a very tedious process. However, we carry it out completely in superfields and then again expand each such supergravity Lagrangian into its component fields. As previously, we ignore  both the Weyl fermion associated with the Galileon supermultiplet as well as the gravitino of supergravity. However, as above, we expand each such Lagrangian to all orders in the Galileon supermultiplet scalar components, $\phi$, $\chi$, and auxiliary field $F$, as well as to all orders in the supergravity multiplet scalars; that is, $g_{\mu\nu}$ with its auxiliary vector field $b_{\mu}$ and complex auxiliary scalar $M$. Having written out the complete expansion into scalar fields, we then show, in detail, that the equations of motion for the supergravity auxiliary fields $b_{\mu}$ and $M$ can be explicitly solved for and present the results. These solutions are then put back into the entire Lagrangian, thus producing the complete $N=1$ supergravitational Lagrangian for the first three conformal Galileons. We have also extended the supersymmetric ${\cal{L}}_{4}$ conformal Galileon to $N=1$ supergravity. However, due to the complexity of the computation, this result will be presented elsewhere. However, we will use several non-trivial results from this work at the end of this paper within the context of low-energy, curved superspace Lagrangians.

Finally, we point out that for the $N=1$ supersymmetric Galileons presented in Section 4 and for their $N=1$ supergravity extensions given in Section 5, we use some of the results and follow the notation presented in \cite{Wess}\footnote{We note, however, that in \cite{Wess} spacetime indices are denoted by Latin letters $m,n,..$. However, to be compatible with much of the literature on higher-derivative supersymmetry and supergravity, in this work we will denote spacetime indices by Greek letters $\mu, \nu, ..$. It will be clear from the context when these refer to spacetime, as opposed to spinorial, indices.}. 

\section{Co-Dimension 1 Brane Action}

\indent In this Section, we review the formalism \cite{Goon:2011qf, Goon:2011uw}  for constructing the worldvolume action of a 3-brane in a 4+1--dimensional bulk space. Denote the bulk space coordinates by
$X^{A}$, $A=0,1,2,3,5$ and the associated metric by $G_{AB}(X)$, where $A=0$ is the time-like direction. The coordinates $X^{A}$ have dimensions of length. We begin by defining a foliation of the bulk space composed of time-like slices. Following \cite{Goon:2011qf, Goon:2011uw}, one chooses coordinates $X^{A}$ so that the leaves of the foliation are the surfaces associated with  $X^{5}=$constant, where the constant runs over a continuous range which depends on the choice of bulk space.
It follows that the  coordinates on an arbitrary leaf of the foliation are  given by  $X^{\mu}$, $\mu=0,1,2,3$. Note that we have denoted the four coordinate indices $A=0,1,2,3$ as $\mu=0,1,2,3$ to indicate that these are the coordinates on the leaves of a time-like foliation.  Now, further restrict the foliation so that it is 1) Gaussian normal with respect to the metric $G_{AB}(X)$ and 2) the extrinsic curvature on each of the leaves of the foliation is proportional to the induced metric. 
Under these circumstances, $X^{5}$ is the transverse normal coordinate and the metric takes the form
\be
G_{AB}(X)dX^{A}dX^{B}=(dX^{5})^{2}+f(X^{5})^{2} g_{\mu\nu}(X) dX^{\mu}dX^{\nu} \ ,
\label{up1}
\ee
where 
$g_{\mu\nu}(X)$ is an arbitrary metric on the foliation and is a function of the four leaf coordinates $X^{\mu}$, $\mu=0,1,2,3$ only. The function $f(X^{5})$ and the intrinsic metric $g_{\mu\nu}(X)$ are dimensionless and will depend on the specific bulk space and foliation geometries of interest. It is important to notice that the coordinates $X^{A}$ satisfying the above conditions and, in particular, the location of their origin, have not been uniquely specified. Although this could be physically important in some contexts, 
for any bulk space of maximal symmetry, such as the $AdS_{5}$ geometry to be discussed in this paper, the origin of such a coordinate system is completely arbitrary and carries no intrinsic information.

\indent Now consider a physical 3+1 brane embedded in the bulk space. Denote a set of intrinsic worldvolume coordinates of the brane by $\sigma^{\mu}$, $\mu=0,1,2,3$. The worldvolume coordinates also have dimensions of length. The location of the brane in the bulk space is specified by the five ``embedding'' functions $X^{A}(\sigma)$ for $A=0,1,2,3,5$, where any given five-tuplet $(X^{(0)}(\sigma), \dots X^{(5)}(\sigma))$ on the brane is a point in the bulk space written in $X^{A}$ coordinates. 
The induced metric and extrinsic curvature on the brane worldvolume are then given by
\be
{\bar{g}}_{\mu\nu}=e^{A}_{~\mu}e^{B}_{~\nu}G_{AB}(X),~~~K_{\mu\nu}=e^{A}_{~\mu}e^{B}_{~\nu}\nabla_{A}n_{B}
\label{rehan}
\ee
where $e^{A}_{~\mu}=\frac{\partial X^{A}}{\partial \sigma^{\mu}}$ are the tangent vectors to the brane and $n_{A}$ is the unit normal vector.\
One expects the worldvolume action to be composed entirely of the geometrical tensors associated with the embedding of the brane into the target space; that is, ${\bar{g}}_{\mu\nu}$ and $K_{\mu\nu}$ defined in \eqref{rehan}, as well as ${\bar{\nabla}}_{\mu}$ and the curvature ${\bar{R}}^{\alpha}_{\beta\mu\nu}$ constructed from ${\bar{g}}_{\mu\nu}$. It follows that the worldvolume action must be of the form
\be
S=\int d^{4}\sigma ~{\cal{L}}\left({\bar{g}}_{\mu\nu}, K_{\mu\nu}, {\bar{\nabla}}_{\mu}, {\bar{R}}^{\alpha}_{\mu\beta\nu} \right) =\int d^{4}\sigma {\sqrt{-{\bar{g}}} {\cal{F}}\left({\bar{g}}_{\mu\nu}, K_{\mu\nu}, {\bar{\nabla}}_{\mu}, {\bar{R}}^{\alpha}_{\mu\beta\nu} \right)} \ ,
\label{3}
\ee
where ${\cal{F}}$ is a scalar function.
\noindent Furthermore, the brane action, and, hence, $\cal{F}$, must be invariant under arbitrary diffeomorphisms of the worldvolume coordinates $\sigma^{\mu}$. Infinitesimal diffeomorphisms are of the form
\be
\delta X^{A}(\sigma)=\xi^{\mu}\partial_{\mu}X^{A}(\sigma)
\label{2}
\ee
for arbitrary local gauge parameters $\xi^{\mu}(\sigma)$. 
Although, naively, there would appear to be five scalar degrees of freedom on the 3-brane worldvolume, it is straightforward to show that one can use the gauge freedom \eqref{2} to set
\be
X^{\mu}(\sigma)=\sigma^{\mu} \ , \quad \mu=0,1,2,3 \ .
\label{4}
\ee
Inverting this expression, it is clear that the worldvolume coordinates $\sigma^{\mu}$ are, in this gauge, fixed to be the bulk coordinates $X^{\mu}$ of the foliation.
The function $X^{5}(\sigma)$, however, is completely unconstrained by this gauge choice. Henceforth, we will always work in the gauge specified by \eqref{4} and define
\be
X^{5}(\sigma) \equiv \pi(\sigma)=\pi(X^{\mu}).
\label{up2}
\ee
That is, there is really only a single scalar function of the transverse foliation coordinates $X^{\mu}$, $\mu=0,1,2,3$ that defines the location of the 3+1 brane relative to the choice of origin of the $X^{A}$ coordinates. We reiterate that, although in some contexts the specific choice of the coordinate origin could be physically important, in a bulk space of maximal symmetry, such as $AdS_{5}$ discussed in this paper, the location of the coordinate origin is completely arbitrary and carries no intrinsic information. 
Note that $\pi(X^{\mu})$ has dimensions of length.

For clarity, let us relate our notation to that which often appears in the literature \cite{Goon:2011qf, Goon:2011uw}. With this in mind, we will denote the four foliation coordinates and the transverse Gaussian normal coordinate by $X^{\mu}\equiv x^{\mu}, \mu=0,1,2,3$ and $X^{5}\equiv\rho$
respectively. It follows that the generic bulk space metric appearing in \eqref{up1} can now be written as
\be
G_{AB}(X)dX^{A}dX^{B}=d\rho^{2}+f(\rho)^{2} g_{\mu\nu}(x) dx^{\mu}dx^{\nu} \ .
\label{mr2}
\ee
Using \eqref{4}and \eqref{up2}, one notes that the scalar field specifying the 3+1 brane location relative to a chosen origin can be expressed as $\rho(x)=\pi(x)$. Therefore, the metric \eqref{mr2} restricted to the brane worldvolume becomes
\be
G_{AB}(X)dX^{A}dX^{B}=d\rho^{2}+f(\pi(x))^{2} g_{\mu\nu}(x) dx^{\mu}dx^{\nu} \ .
\label{mr3}
\ee
It then follows that the induced metric and the extrinsic curvature on the  brane are given by
\be
{\bar{g}}_{\mu\nu}=f(\pi)^{2}g_{\mu\nu}+{\nabla}_{\mu} \pi {\nabla}_{\nu} \pi , \qquad
K_{\mu\nu}=\gamma \Big(-\nabla_{\mu}\nabla_{\nu}\pi +f{f'}g_{\mu\nu}+2\frac{{f'}}{f} \nabla_{\mu} \pi\nabla_{\nu}\pi \Big)
\label{6}
\ee
respectively, where ${'=\partial/\partial \pi}$ and 
\be
\gamma=\frac{1}{\sqrt{1+\frac{1}{f^{2}}(\nabla \pi)^{2}}} \ .
\label{7}
\ee

\indent An action of the form \eqref{3} will generically lead to equations of motion for the physical scalar field $\pi(x)$ that are higher than second order in derivatives and, hence, possibly propagate extra ghost degrees of freedom. Remarkably, this can be avoided \cite{deRham:2010eu, Goon:2011qf, Goon:2011uw} if one restricts the Lagrangian to be of the form
\be
{\cal{L}}=\Sigma_{i=1}^{5} ~c_{i} {\cal{L}}_{i} ,
\label{8}
\ee
where the $c_{i}$ are constant real coefficients,
\bea
{\cal{L}}_1 &=& \sqrt{-g}\int^{\pi}d\pi' f(\pi')^{4},
\nn \\
{\cal{L}}_2 &=& -\sqrt{-{\bar{g}}},
\nn \\
{\cal{L}}_3 &=& \sqrt{-{\bar{g}}} ~K,
\nn \\
{\cal{L}}_4 &=& -\sqrt{-{\bar{g}}} ~ {\bar{R}},
\nn \\
{\cal{L}}_5 &=& \frac{3}{2}\sqrt{-{\bar{g}}}~{K}_{GB}
\label{9}
\eea
with $K={\bar{g}}^{\mu\nu}K_{\mu\nu}$, ${\bar{R}}={\bar{g}}^{\mu\nu}{\bar{R}}^{\alpha}_{\mu\alpha\nu}$ and ${\cal{K}}_{GB}$ is a Gauss-Bonnet boundary term given by
\be
{\cal{K}}_{GB}= -\frac{1}{3}K^{3} +K_{\mu\nu}^{2}K- \frac{2}{3} K_{\mu\nu}^{3}-2\Big( {\bar{R}}_{\mu\nu}-\frac{1}{2}{\bar{R}}{\bar{g}}_{\mu\nu} \Big) K^{\mu\nu} \ .
\label{10}
\ee
All indices are raised and traces taken with respect to ${\bar{g}}^{\mu\nu}$. It has been shown \cite{deRham:2010eu, Goon:2011qf, Goon:2011uw} that Lagrangian \eqref{8}, for any choices of coefficients $c_{i}$,  leads to an equation of motion for $\pi(X^{\mu})$ that is only second order in derivatives. In this paper, we will assume that both \eqref{3} and \eqref{8},\eqref{9} are satisfied.
 
\indent Evaluating each of the Lagrangians in \eqref{9} for an arbitrary metric of the form \eqref{mr2} is arduous and has been carried out in several papers \cite{deRham:2010eu, Goon:2011qf, Goon:2011uw}. The ${\cal{L}}_{5}$ term is particularly long and not necessary for the work to be discussed here. Hence, we will ignore it in the rest of this paper. The remaining four Lagrangians are found to be
\bea
{\cal{L}}_1 &=&  \sqrt{-g} \int^{\pi} d\pi' f^4(\pi') 
\nn \\
\nn \\
{\cal{L}}_2 &=& - \sqrt{-g} f^4 \sqrt{1 + \frac{1}{f^{2}} (\nabla \pi)^2}
\nn \\
\nn \\
{\cal{L}}_3 &=&  \sqrt{-g} \bigg[ 
f^3 f' (5 - \gamma^2) - f^2 [\Pi] + \gamma^2 [\pi^3] \bigg]
\nn \\
\nn \\
{\cal{L}}_4 &=& -\sqrt{-g} \bigg\{ \frac{1}{\gamma}f^{2}R -2 \gamma R_{\mu\nu} \nabla^{\mu}\pi \nabla^{\nu} \pi+
\gamma \bigg[
[\Pi]^2 - [\Pi^2] + 2 \gamma^2 \frac{1}{f^{2}} \big(  - [\Pi] [\pi^3] + [\pi^4]\big)
\bigg] 
\nn \\
&&+ 6 \frac{f^3 f''}{ \gamma} (-1 + \gamma^2)+  2\gamma f f' \bigg[ -4 [\Pi] + \frac{\gamma^2}{ f^{2}} \big( f^2 [\Pi] + 4 [\pi^3]\big)\bigg]
\nn \\
&&-6\frac{f^2 (f')^2}{\gamma} (1- 2\gamma^2 + \gamma^4) \bigg \}
\label{11}
\eea
In these expressions, all covariant derivatives and curvatures are with respect to the foliation metric $g_{\mu\nu}$.
We follow the notation common in the literature. Defining $\Pi_{\mu \nu} \equiv \nabla_\mu \nabla_\nu \pi$, the bracket $[\Pi^{n}]$ denotes the trace of n-powers of $[\Pi]$ with respect to $g^{\mu \nu}$. For example, $[\Pi]=\nabla_\mu \nabla^\mu \pi$, $[\Pi^{2}]=\nabla_\mu \nabla_\nu \pi \nabla^\mu \nabla^\nu \pi$ and so on. Similarly, we also define contractions of powers of $\Pi$ with $\nabla \pi$ using the notation $[\pi^{n}] \equiv \nabla \pi \Pi^{n-2} \nabla \pi$. For example, $[\pi^{2}]=\nabla_{\mu}\pi \nabla^{\mu} \pi$, $[\pi^{3}]=\nabla_{\mu} \pi \nabla^{\mu} \nabla^{\nu} \pi \nabla_{\nu}\pi$ and so on. Note that the Lagrangians ${\cal{L}}_1,  {\cal{L}}_2,  {\cal{L}}_3$ and ${\cal{L}}_4$  in \eqref{11} have mass dimensions $-1, 0, 1$ and $2$ respectively. Hence, the constant coefficients  $c_{1}, c_{2},c_{3}, c_{4}$ in action \eqref{8} have mass dimensions $5, 4, 3$ and $2$.

\section{A Flat 3-Brane in $AdS_{5}$: Conformal Galileons}

Henceforth, we will restrict our discussion to the case where the target space is the ``maximally symmetric'' 5-dimensional anti-de Sitter space $AdS_{5}$ with isometry algebra $so(4,2)$ and the foliation leaves are ``flat''--that is, have Poincare isometry algebra $p(3,1)$
\footnote{The case of the 5-dimensional Poincare space, leading to the ``Poincare" Galileons, and their extension to supersymmetry has been discussed in \cite{Farakos:2013zya}.}. This geometry is easily shown to satisfy the above two assumptions that the foliations are Gaussian normal with respect to the target space metric and the extrinsic curvature is proportional to the induced metric. It then follows that the $AdS_{5}$ metric written in the $X^{A}$ coordinates subject to gauge choice  \eqref{2} and definition \eqref{up2} is of the form \eqref{mr3}. More specifically, if we denote the $AdS_{5}$ radius of curvature by ${\cal{R}}(>0)$, and denote the flat metric on the foliations by $\eta_{\mu\nu}$, one finds that the target space metric is given by
\be
G_{AB}dX^{A}dX^{B}=d\rho^{2}+f(\pi)^{2}\eta_{\mu\nu}dx^{\mu}dx^{\nu} ,
\label{12}
\ee
where
\be
f(\pi)=e^{-\frac{\pi}{\cal{R}}} .
\label{13}
\ee
It follows that the four Lagrangians given in \eqref{11} become
\bea
{\cal{L}}_1 &=&-\frac{{\cal{R}}}{4} e^{-\frac{4\pi}{\cal{R}}}
\nn \\
\nn \\
{\cal{L}}_2 &=& - e^{-\frac{4\pi}{\cal{R}}} \sqrt{1 + e^{\frac{2\pi}{\cal{R}}}(\pt \pi)^2}
\nn \\
\nn \\
{\cal{L}}_3 &=&  \gamma^2 [\pi^3] -  e^{-\frac{2\pi}{\cal{R}}}[\Pi] + \frac{1}{{\cal{R}}} e^{-\frac{4\pi}{\cal{R}}}(\gamma^{2}-5) 
\nn \\
\nn \\
{\cal{L}}_4 &=& -\gamma ([\Pi]^2 - [\Pi^2]) -2\gamma^{3}e^{\frac{2\pi}{\cal{R}}}([\pi^4]- [\Pi] [\pi^3])
\nn \\
&&+ \frac{6}{{\cal{R}}^{2}} e^{-\frac{4\pi}{\cal{R}}} \frac{1}{ \gamma} (2 -3 \gamma^2+\gamma^{4})+ \frac{8}{\cal{R}} \gamma^{3} [\pi^{3}]- \frac{2}{\cal{R}}e^{-\frac{2\pi}{\cal{R}}}
\gamma (4-\gamma^{2})[\Pi] 
\label{14}
\eea
respectively, where 
\be
\gamma=\frac{1}{\sqrt{1+e^{\frac{2\pi}{\cal{R}}}(\pt \pi)^{2}}} 
\label{15}
\ee
and $[\Pi^{n}]$, $[\pi^{n}]$ are defined as above with $\nabla \rightarrow \pt$. These are precisely the conformal DBI Galileons, first presented in \cite{deRham:2010eu, Goon:2011qf, Goon:2011uw}.
It can be shown that each of the terms in \eqref{14} is invariant, up to a total divergence, under the transformations
\be
\delta \pi= {\cal{R}}-x^{\mu}\pt_{\mu}\pi~,\qquad \delta_{\mu}\pi= 2x_{\mu}+({\cal{R}} e^{\frac{2\pi}{\cal{R}}}+ \frac{1}{{\cal{R}}}x^{2})\pt_{\mu}\pi - \frac{2}{{\cal{R}}}x_{\mu}x^{\nu}\pt_{\nu}\pi \ .
\label{16}
\ee

\indent  Defining the dimensionless field and the $AdS_{5}$ mass scale by
\be
\hat{\pi}=\frac{\pi}{{\cal{R}}} \ , \qquad  {\cal{M}}=1 / {\cal{R}} 
\label{17}
\ee
respectively, it is clear that each of the four conformal DBI Lagrangians in \eqref{14} admits an expansion in powers of $(\frac{\pt}{{\cal{M}}})^{2}$. Performing this expansion and combining terms with the same power of $(\frac{\pt}{{\cal{M}}})^{2}$ arising in different Lagrangians \eqref{14}, one can, up to total derivatives,  re-express the action ${\cal{L}}=\Sigma_{i=1}^{4}c_{i}{\cal{L}}_{i}$ as
\be
{\cal{L}}=\Sigma_{i=1}^{4} {\bar{c}}_{i}{\bar{\cal{L}}}_{i}
\label{burt1}
\ee
where
\bea
{\bar{c}}_{1} &=&  \frac{c_{1}}{\cal{M}}+4c_{2}+16{\cal{M}}c_{3},
\nn \\
{\bar{c}}_{2} &=&  \frac{c_{2}}{{\cal{M}}^{2}}+6\frac{c_{3}}{\cal{M}}+12c_{4},
\nn \\
{\bar{c}}_{3} &=& \frac{c_{3}}{{\cal{M}}^{3}}+6\frac{c_{4}}{{\cal{M}}^{2}},
\nn \\
{\bar{c}}_{4} &=&\frac{c_{4}}{ {\cal{M}}^{4}}
\label{burt2}
\eea
are real constants and
\bea
{\bar{\cal{L}}}_1 &=& -\frac{1}{4} e^{-4{\hat{\pi}}}
\nn \\
\nn \\
{\bar{\cal{L}}}_2 &=& - \frac{1}{2}e^{-2{\hat{\pi}}} (\frac{\pt {\hat{\pi}}}{{\cal{M}}})^{2}
\nn \\
\nn \\
{\bar{\cal{L}}}_3 &=&  \frac{1}{2}(\frac{\pt {\hat{\pi}}}{{\cal{M}}})^{2}
\frac{\Box{\hat{\pi}}}{{\cal{M}}^{2}} -\frac{1}{4}(\frac{\pt {\hat{\pi}}}{{\cal{M}}})^{4}
\nn \\
\nn \\
{\bar{\cal{L}}}_4 &=& e^{-2{\hat{\pi}}} (\frac{\pt {\hat{\pi}}}{{\cal{M}}})^{2} \big[-\frac{1}{2}(\frac{\Box {\hat{\pi}}}{{\cal{M}}^{2}})^{2}  +\frac{1}{2} (\frac{\pt_{\mu}\pt_{\nu}{\hat{\pi}} }{{\cal{M}}^{2}})( \frac{\pt^{\mu}\pt^{\nu}{\hat{\pi}} }{{\cal{M}}^{2}})
\nn \\
&& \qquad \qquad\qquad\qquad\quad~  -\frac{1}{5} (\frac{\pt {\hat{\pi}}}{{\cal{M}}})^{2}\frac{\Box {\hat{\pi}}}{{\cal{M}}^{2}}  
+ \frac{1}{5}(\frac{\pt_{\mu} {\hat{\pi}}}{{\cal{M}}})(\frac{\pt_{\nu} {\hat{\pi}}}{{\cal{M}}}) ( \frac{\pt^{\mu}\pt^{\nu}{\hat{\pi}} }{{\cal{M}}^{2}}) -\frac{3}{20}(\frac{\pt {\hat{\pi}}}{{\cal{M}}})^{4}  \big]
\label{18}
\eea
We have chosen each Lagrangian in \eqref{18} to be dimensionless and, hence, each coefficient ${\bar{c}}_{i}$ has dimension 4. Note that \eqref{18} are precisely the first four conformal Galileons. Since the original coefficients $c_{i},~i=1\dots 4$ are arbitrary, it follows from \eqref{burt2} that the coefficients ${\bar{c}}_{i},~i=1\dots 4$ are also unconstrained. We find from \eqref{16} and \eqref{17} that, in this expansion, each Lagrangian in \eqref{18} is invariant under the conformal Galileon symmetry
\be
\delta {\hat{\pi}}=1-x^{\mu} \pt_{\mu} {\hat{\pi}}, \qquad \delta_{\mu}{\hat{\pi}}=2x_{\mu} + x^{2} \pt_{\mu} {\hat{\pi}}-2x_{\mu}x^{\nu}\pt_{\nu}{\hat{\pi}} \ .
\label{19}
\ee
\indent We conclude that, expanded up to sixth-order in $(\partial/\cal{M})$, the worldvolume Lagrangian for a flat 3-brane embedded in $AdS_{5}$ is given by
\be
{\cal{L}}=\Sigma_{i=1}^{4} {\bar{c}}_{i} {\bar{\cal{L}}}_{i} \ ,
\label{20}
\ee
where each Lagrangian ${\bar{\cal{L}}}_{i}$ and each constant coefficient ${\bar{c}}_{i}$ have mass dimensions 0 and 4 respectively. As discussed previously, we are, for simplicity, ignoring the fifth Galileon  which is eighth-order in $(\partial/\cal{M})$--since it is not necessary in this paper. However, it can easily be included without changing any of our results. Note that all terms of order greater than 8 in the derivative expansion of the DBI conformal Galileons can be shown to be a total divergence \cite{Nicolis:2008in,Creminelli:2013ygt} and, hence, do not contribute to the theory.\footnote {A more complete discussion of the  $(\frac{\pt}{{\cal{M}}})^{2}$ expansion is the following. Unlike the discussion in this Section, let us here include the Lagrangian ${\cal{L}}_{5}$ in the sum ${\cal{L}}=\Sigma_{i=1}^{5} ~c_{i} {\cal{L}}_{i}$ as in \eqref{8}. Now perform the  derivative expansion of the ${\cal{L}}_{i}$ for $i=1\dots 5$ to all orders in $(\frac{\pt}{{\cal{M}}})^{2}$. It is well-known \cite{Nicolis:2008in,Creminelli:2013ygt} that all terms with $(\frac{\pt}{{\cal{M}}})^{2p}$ for $p>4$ form a total divergence and, hence, can be ignored in the action. Therefore, this expansion is exact and does not require that one demand that $(\frac{\pt}{{\cal{M}}})^{2} \ll 1$. This is unique to the case of the conformal Galileons that we are discussing.}

\section{Supersymmetric Conformal Galileons}

In a previous paper \cite{Khoury:2011da}, the real scalar field ${\hat{\pi}}$ and Lagrangians ${\bar{\cal{L}}}_{i}$, $i=2\dots 5$ were extended to flat space  $N=1$ supersymmetry. To do this, it was convenient to define a dimensionless real scalar field
\be
\phi \equiv e^{{\hat{\pi}}}
\label{21}
\ee
and set ${\cal{M}}=1$. Here, we will review this analysis, again neglecting  ${\bar{\cal{L}}}_{5}$, but with one important new ingredient. That is, we now include the Lagrangian ${\bar{{\cal{L}}}}_{1}$ given in \eqref{18}. This adds a potential energy term to the scalar Lagrangian and, hence, requires a non-vanishing superpotential to appear in the superfield action. In turn, this necessitates a more subtle discussion of the auxiliary $F$-field which occurs in the component field expansion of the super-Lagrangian. In particular, we give a careful analysis of how it can be eliminated via its equation of motion and what constraints, if any, that puts on the coefficients ${\bar{c}}_{i}$. In this section, as well as in the next section on supergravity, we use results and follow the notation presented in \cite{Wess}

\indent We begin by presenting  ${\bar{\cal{L}}}_{i}$, $i=1\dots4$ in \eqref{18} in terms of the $\phi$ field defined in \eqref{21} with ${\cal{M}}$ set to unity. The result is
\bea
{\bar{{\cal L}}}_1 &=& -\frac{1}{4 \p^4} \nn \\
{\bar{{\cal L}}}_2 &=& -\frac{1}{2\p^4}(\pt \p)^2 \nn \\
{\bar{{\cal L}}}_3 &=& \frac{1}{2\p^3}\Box\p(\pt\p)^2 -\frac{3}{4\p^4}(\pt\p)^4\nn \\ 
{\bar{{\cal L}}}_4 &=& -\frac{1}{2\p^2}(\pt\p)^2(\Box\p)^2 +\frac{1}{2\p^2}(\pt\p)^2\p^{,\mu\nu}\p_{,\mu\nu} +\frac{4}{5\p^3}(\pt\p)^4\Box\p\nn \\
&& -\frac{4}{5\p^3}(\pt\p)^2\p^{,\m}\p^{,\nu}\p_{,\mu\nu}-\frac{3}{20\p^4}(\pt\p)^6  \nn \\
&=& -\frac{1}{4\p^2}\pt_\mu(\pt\p)^2\pt^\mu(\pt\p)^2 +\frac{1}{\p^2}\Box\p\p^{,\mu}\p^{,\nu}\p_{,\mu\nu} -\frac{1}{4\p^3}(\pt\p)^4\Box\p\,.
\label{22}
\eea
where the second versions of ${\bar{{\cal L}}}_4$ follows from integration by parts. 

\indent Having presented the Galileon Lagrangians associated with the real scalar field $\phi$,\footnote{We have written the conformal Galileons in \eqref{18} in terms of the field $\phi$ defined in \eqref{21} so as to greatly simplify the extension to the supersymmetric case. By doing so, the $-\infty < {\hat{\pi}} < +\infty$ range of the field ${\hat{\pi}}$ is changed to the $0< \phi <+\infty$ regime of field $\phi$. Of course, the $\phi \rightarrow 0$ surface is equivalent to ${\hat{\pi}} \rightarrow -\infty$ and, hence, $\phi$ can only approach zero, but never achieve it. Hence, nowhere in the range of $\phi$ do the Lagrangians, or any other quantity in our derivation, diverge. It is equally possible to work directly with the $\hat{\pi}$ field, but the supersymmetrization, althought completely equivalent to that presented in this paper, is far more complicated--both to construct and as mathematical expressions.} we now embed $\phi$ in an $N=1$ chiral superfield
\be
\Phi=(A,\psi,F) \ .
\label{23}
\ee
Here $A=\frac{1}{\sqrt{2}}(\phi+i\chi)$ is a complex field composed of two real scalar fields $\phi$ and $\chi$, $\psi$ is a two-component Weyl spinor and $F$ is a complex ``auxiliary''
field that can, for Lagrangians with at most two derivatives on the scalar fields, be eliminated from the super-Lagrangian using its equation of motion. Note that since scalars $\phi$ and, hence, $\chi$ are dimensionless, and since the anticommuting superspace coordinate $\theta$ has mass dimension $-1/2$, then the complex scalar A, the Weyl spinor $\psi$ and the complex scalar $F$ have dimensions $0$, $1/2$ and $1$ respectively. The role of the $F$-field in higher-derivative Lagrangians without a potential energy was discussed in \cite{Khoury:2011da}. In the present paper, however, we will carefully re-examined the $F$-field, this time in the presence of a non-vanishing potential. Ignoring ${\bar{\cal{L}}}_{1}$, the supersymmetric extensions of ${\bar{\cal{L}}}_{2}$, ${\bar{\cal{L}}}_{3}$ and ${\bar{\cal{L}}}_{4}$ were constructed in \cite{Khoury:2011da}, both in superfields and in their component field expansion--working, however, only to quadratic order in all component fields except $\phi$. In this paper, we present the same superfield expressions as in \cite{Khoury:2011da}. However, unlike that paper, we will not display any component field terms containing the fermion--since this is not of interest in this work. On the other hand, we give the full component field expansion for all scalars $\phi, \chi$ and $F$, since this will be important for our discussion of the equation for $F$. The results are the following.

\subsection{${\bar{{\cal{L}}}}_{2}$}

Defining
\begin {equation}
K(\Phi,{\Phi}^{\dagger})=\frac{2}{3(\Phi+\Phi^{\dagger})^{2}} \ ,
\label{24}
\end{equation}
the complete supersymmetrized ${\bar{{\cal{L}}}}_{2}$ action is given by
\bea
{\bar{{\cal{L}}}}_{2}^{\rm SUSY}&=&K(\Phi,{\Phi}^{\dagger})\Big\vert_{\theta \theta {\bar{\theta}} {\bar{\theta}}}\nn \\
&=&-\frac{1}{2\phi^{4}}(\pt\phi)^{2}-\frac{1}{2\phi^{4}}(\pt\chi)^{2}+\frac{1}{\phi^{4}}F^{*}F  \ .
 \label{25}
\eea
Note that this matches the corresponding expression in \eqref{22} when $\chi=F=0$. For specificity, it is useful to reintroduce the mass ${\cal{M}}=1/{\cal{R}}$ into \eqref{25}. This dimensionless Lagrangian then becomes
\bea
{\bar{{\cal{L}}}}_{2}^{\rm SUSY}=-\frac{1}{2\phi^{4}}(\frac{\pt\phi}{{\cal{M}}})^{2}-\frac{1}{2\phi^{4}}(\frac{\pt\chi}{\cal{M}})^{2}+\frac{1}{\phi^{4}}\frac{F^{*}F}{{\cal{M}}^{2}}  \ .
 \label{25A}
\eea
It follows that the symbols $\partial$ and $F$ in \eqref{25} stand for $\frac{\partial}{\cal{M}}$ and $\frac{F}{\cal{M}}$ respectively, where we have, for simplicity, set ${\cal{M}}=1$. This will be the case for the remainder of this paper, unless otherwise specified. Finally, we find it convenient to re-express \eqref{25} in terms of the complex scalar field  $A=\frac{1}{\sqrt{2}}(\phi+i\chi)$ and the lowest component of the Kahler potential defined in \eqref{24}. We find that
\be
{\bar{{\cal{L}}}}_{2}^{\rm SUSY}=- \frac{\partial^2 K}{\partial A \partial A^*} \partial A \cdot \partial A^*
+  \frac{\partial^2 K}{\partial A \partial A^*} F F^*
\label{bw1}
\ee

\subsection{${\bar{{\cal L}}}_3$}

The complete supersymmetrized ${\bar{{\cal{L}}}}_{3}$ action is given as a specific sum of two superfield Lagrangians. These are 
\bea
\bigg[\frac{1}{(\Phi + \Phi^\dagger)^3} \bigg( D \Phi D \Phi \bar{D}^2 \Phi^\dagger + \mathrm{h.c.} \bigg) \bigg] \bigg|_{\theta\theta\bar{\theta}\bar{\theta}}
&=&\frac{4}{\phi^3}(\partial \phi)^2 \square \phi - \frac{4}{\phi^3}(\partial \chi)^2 \square \phi  + \frac{8}{\phi^3} (\partial \phi \cdot \partial \chi) \square \phi
\nn \\
&-& \big( \frac{8}{\phi^3} \square \phi 
+ \frac{12}{\phi^4} (\partial \phi)^2 + \frac{12}{\phi^4} (\partial \chi)^2\big) F^* F \nn \\
&+& \frac{8i}{\phi^3} \chi_{, \mu} (F^* \partial^{\mu}F - F \partial^{\mu} F^{*})+ \frac{24}{\phi^4} (F^* F)^2 \
 \label{26}
 \eea
and
\begin{eqnarray}
\bigg[ \frac{1}{(\Phi + \Phi^\dagger)^4} D \Phi D \Phi \bar{D} \Phi^\dagger \bar{D} \Phi^\dagger \bigg] \bigg|_{\theta\theta\bar{\theta}\bar{\theta}}
&=&
\frac{1}{\phi^4}(\partial \phi)^4 + \frac{1}{\phi^4}(\partial \chi)^4 - \frac{2}{\phi^4}(\partial \phi)^2 (\partial \chi)^2 + \frac{4}{\phi^4}(\partial \phi \cdot \partial \chi)^2
\nn \\
&-& \big( \frac{4}{\phi^4}(\partial \phi)^2 + \frac{4}{\phi^4}(\partial \chi)^2 \big) F^* F +\frac{4}{\phi^4}(F^* F)^2  
\label{27}
\end{eqnarray}
respectively. Note that, as discussed above, we have dropped all terms containing the fermion but have included all of the scalar fields to all orders. These can be combined to give a supersymmetric extension of the ${\bar{{\cal L}}}_3$ conformal Galileon Lagrangian
\bea
\bar{\mathcal{L}}_{3}^{\mathrm{SUSY}}
&=& 
\frac{1}{8}\bigg[\frac{1}{(\Phi + \Phi^\dagger)^3} \bigg( D \Phi D \Phi \bar{D}^2 \Phi^\dagger + \mathrm{h.c.} \bigg) \bigg] \bigg|_{\theta\theta\bar{\theta}\bar{\theta}}
-\frac{3}{4}
\bigg[ \frac{1}{(\Phi + \Phi^\dagger)^4} D \Phi D \Phi \bar{D} \Phi^\dagger \bar{D} \Phi^\dagger \bigg] \bigg|_{\theta\theta\bar{\theta}\bar{\theta}}
\nn \\
&=&
\frac{1}{2\phi^3} (\pa \phi)^2 \square \phi - \frac{3}{4\phi^4} (\pa \phi)^4- \frac{1}{2\phi^3} (\pa \chi)^2 \square \phi- \frac{3}{4\phi^4} (\pa \chi)^4  
\nn \\
&+& \frac{1}{\phi^3} (\pa \phi \cdot \pa \chi)\square \phi 
+ \frac{3}{2\phi^4}(\pa \phi)^2 (\pa \chi)^2 - \frac{3}{\phi^4}(\pa \phi \cdot \pa \chi)^2 
\nn \\
&+& \big( -\frac{1}{\phi^3} \square \phi + \frac{3}{2\phi^4}(\pa \phi)^2  + \frac{3}{2\phi^4}(\pa \chi)^2\big)F^* F
 + \frac{i}{\phi^3} \chi_{, \mu} (F^* \partial^{\mu}F - F \partial^{\mu}F^*) \ .
\label{28}
\eea
Note that~(\ref{28}) reduces to ${\bar{{\cal L}}}_{3}$ in \eqref{22} when $\chi=F=0$, as it should. Furthermore, it is important to note that the quartic term $(F^*F)^{2}$ has cancelled between the first and second expressions, \eqref{26} and \eqref{27} respectively. This will play an important role in our discussion of the $F$ equation. Again, it is useful to re-express \eqref{28} in terms of the complex scalar field $A$ and the lowest component of the Kahler potential $K$. The result is
\bea
\bar{\mathcal{L}}_{3}^{\mathrm{SUSY}}&=& \frac{1}{(A+ A^*)^3}\bigg[
2 \big(\partial A \big)^2 \square A^*
+ 2 \big(\partial A^* \big)^2 \square A
- 2 F^* F \square ( A + A^*)
\nn \\
&&\qquad \qquad \qquad
+ 2 F^* \partial F \cdot \partial (A - A^*)
- 2 F \partial F^* \cdot \partial (A - A^*)
\bigg] 
\nn \\
&-&
\frac{6}{(A+ A^*)^4}
\bigg[
2(\partial A)^2 (\partial A^*)^2
+  6 (\partial A \cdot \partial A^*) F F^*
\bigg] \ .
\label{bw2}
\eea

\subsection{${\bar{{\cal L}}}_4$}

We now supersymmetrize ${\bar{{\cal L}}}_4$. It is convenient to use the second expression for ${\bar{{\cal L}}}_4$ in \eqref{22}, obtained using integrating by parts. This expression is simpler, consisting of only three terms. We proceed by first constructing the  supersymmetric extension for each of these terms. For the first term, consider
\begin{eqnarray}
\bar{\mathcal{L}}_{4, \, \mathrm{1st\,term}}^{\mathrm{SUSY}}
&=&
\frac{1}{64(\Phi + \Phi^\dagger)^2} \{ D, \bar{D}\} (D\Phi D\Phi) \{ D, \bar{D} \} (\bar{D} \Phi^\dagger \bar{D} \Phi^\dagger)\bigg|_{\theta\theta\bar{\theta}\bar{\theta}}
\nn \\
&=&
\frac{1}{4\phi^2} \pa_\mu (\pa \phi)^2 \pa^\mu (\pa \phi)^2 
\nn \\
&-& \frac{1}{2\phi^2} \pa_\mu (\pa \phi)^2 \pa^\mu (\pa \chi)^2 
+ \frac{1}{2\phi^2} \pa_\mu (\pa \phi \cdot \pa \chi) \pa^\mu (\pa \phi \cdot \pa \chi) 
+ \frac{1}{4\phi^2} \pa_\mu (\pa \chi)^2 \pa^\mu (\pa \chi)^2
\nn \\
&-& \frac{1}{\phi^2} \big( \phi_{,\mu \nu} \phi^{, \mu \nu} + \chi_{,\mu \nu} \chi^{, \mu \nu} \big) F F^* 
- \frac{1}{\phi^2} \big( \phi_{,\mu} \phi^{,\mu \nu}  + \chi_{,\mu} \chi^{,\mu \nu}  \big)
 \big( F^*\partial_{\mu}F  +  F\partial_{\mu}F^* \big)
\nn \\
&+& \frac{1}{\phi^2}\big( (\pa \phi)^2  + (\pa \chi)^2 \big) \pa F^* \cdot \pa F
+ \frac{4}{\phi^2} F^* F \big(\pa F^* \cdot \pa F \big)
\label{29}
\end{eqnarray}
where, in component fields, we have dropped all terms containing the fermion, but work to all orders in the scalar fields $\p$, $\chi$ and $F$ and used integration by parts.
Note that this reduces to the first term for ${\bar{{\cal L}}}_4$ in \eqref{22} when $\chi=F=0$. The second term can be supersymmetrized as
\bea
\nn
{\bar{\cal L}}_{4,\,{\rm 2nd}\;{\rm term}}^{\rm SUSY}  &=& \frac{-1}{128(\P +\Pd)^2}\left(\{ D,\Db \}(\P + \Pd) \{ D,\Db \}( D\P D\P) \Db^2 \Pd +{\rm h.c.}\right) \bigg\vert_{\th\th\tb\tb} \\
&=& 
-\frac{1}{2\phi^2} \phi^{,\mu} \pa_{\mu} (\pa \phi)^2 \square \phi
\nn \\
&&
+ \frac{1}{2\phi^2} \phi^{,\mu} \pa_{\mu} (\pa \chi)^2 \square \phi 
+ \frac{1}{\phi^2} \phi^{,\mu} \pa_{\mu} (\pa \phi \cdot \pa \chi) \square \chi
\nn \\
&&
+ \bigg(
\frac{1}{\phi^3} \phi^{,\mu}\pa_\mu \big(  (\pa \phi)^2  -  (\pa \chi)^2 \big) 
+ \frac{2}{\phi^3} \chi_{,\mu} \chi^{,\mu \nu} \phi_{,\nu}
- \frac{2}{\phi^2} \chi_{, \mu \nu} \chi^{, \mu \nu}
+\frac{1}{\phi^2} \phi^{,\mu} \square \big( \phi_{,\mu} + \chi_{,\mu}\big)
\bigg)F F^*
\nn \\
\nn \\
&&
+ \bigg( 
\frac{1}{\phi^3}\phi^{,\mu}(\pa \chi)^2 
- \frac{i}{\phi^3} \phi^{,\mu}(\pa \phi \cdot \pa \chi)
+\frac{1}{2\phi^2} \phi^{,\mu} \square ( \phi + \chi ) 
\nn \\
&& 
- \frac{1}{4\phi^2} \big( (\pa \phi)^2 - (\pa \chi)^2 + 2i \pa \phi \cdot \pa \chi \big)^{,\mu}
+ \frac{1}{2\phi^2} \chi^{,\mu \nu} \big(  \chi_{,\nu} - i \phi_{,\nu} \big)
\nn \\
&&
+\frac{1}{2\phi^2} \phi_{,\nu}\big( \phi^{,\mu \nu} - i \chi^{,\mu \nu}\big)
\bigg) F^*\partial_{\mu}F 
\nn \\
\nn \\
&&
+ \bigg(
\frac{1}{\phi^3}\phi^{,\mu}(\pa \chi)^2 
- \frac{i}{\phi^3} \phi^{,\mu}(\pa \phi \cdot \pa \chi)
+\frac{1}{2\phi^2} \phi^{,\mu} \square ( \phi + \chi )  
\label{30}
\\
&& 
- \frac{1}{4\phi^2} \big( (\pa \phi)^2 - (\pa \chi)^2 + 2i \pa \phi \cdot \pa \chi \big)^{,\mu}
+ \frac{1}{2\phi^2} \chi^{,\mu \nu} \big(  \chi_{,\nu} + i \phi_{,\nu} \big)
\nn \\
\nn \\
&& 
+\frac{1}{2\phi^2} \phi_{,\nu}\big( \phi^{,\mu \nu} + i \chi^{,\mu \nu}\big)
\bigg)F \partial_{\mu}F^*
\nonumber \\
&&
+\frac{1}{\phi^2} \phi^{,\mu} \phi^{,\nu} F_{,\mu} F_{\nu}^* 
+ \frac{i}{2\phi^2}\big(
\chi^{,\mu } \phi^{,\nu}  
- \chi^{,\nu } \phi^{,\mu}  
\big)F_{,\mu}^* F_{\nu}
+ \frac{2}{\phi^2} F^* F \pa^{\mu} F^* \pa_{\mu} F
\nn \\
&&
+ \frac{1}{\phi^3} F (F^*)^2 (\pa F \cdot \pa \phi)
+ \frac{1}{\phi^3} F^* F^2 (\pa F^* \cdot \pa \phi)\ .
\nn \\
\nonumber
\eea
When $\chi=F=0$, this is simply the second term for ${\bar{{\cal L}}}_4$ in \eqref{22}. 
Finally, consider the third term. As discussed in \cite{Khoury:2011da}, there are two inequivalent ways of supersymmetrizing this term. For simplicity, we will focus on the easiest such supersymmetrization. This is given by
\bea
\bar{{\cal L}}_{4,\,{\rm 3rd ~term}}^{\rm SUSY} &=& \frac{1}{64(\P+\Pd)^3} D\P D\P\Db\Pd\Db\Pd\{ D,\Db \} \{ D,\Db \}(\P + \Pd)  \Big\vert_{\th\th\tb\tb}  \nn\\ 
&=&  
\frac{1}{4\phi^3} (\pa \phi)^4 \square \phi
\nn \\
&&
+ \frac{1}{4 \phi^3} (\pa \chi)^4 \square \phi 
- \frac{1}{2\phi^3}(\pa \phi)^2 (\pa \chi)^2 \square \phi 
+ \frac{1}{\phi^3}(\pa \phi \cdot \pa \chi)^2 \square \phi
\nn \\
&&
- \frac{1}{\phi^3} \bigg( (\pa \phi)^2 \square \phi + (\pa \chi)^2 \square \phi \bigg) F F^*
\nn \\
&&
+ \frac{1}{\phi^3} (F F^*)^2 \square \phi \ . 
\label{31}
\eea
For $\chi=F=0$, this gives the third term for ${\bar{{\cal L}}}_4$ in \eqref{22}. Note that in both \eqref{30} and \eqref{31}, the component field expressions have been obtained by dropping all terms containing the fermion, but, as in the first term, working to all orders in the scalar fields $\p$, $\chi$ and $F$.

Putting these three terms together, we get a complete supersymmetrization of ${\bar{{\cal L}}}_4$ in \eqref{22}. This is given by
\begin{eqnarray}
{\bar{\mathcal{L}}}_{4}^{\mathrm{SUSY}}&=& \frac{1}{64(\Phi + \Phi^\dagger)^2} \{ D, \bar{D}\} (D\Phi D\Phi) \{ D, \bar{D} \} (\bar{D} \Phi^\dagger \bar{D} \Phi^\dagger)\bigg|_{\theta\theta\bar{\theta}\bar{\theta}}
\nn \\
&+& \frac{-1}{128(\P +\Pd)^2}\left(\{ D,\Db \}(\P + \Pd) \{ D,\Db \}( D\P D\P) \Db^2 \Pd +{\rm h.c.}\right) \bigg\vert_{\th\th\tb\tb} \nn \\
&+&  \frac{1}{64(\P+\Pd)^3} D\P D\P\Db\Pd\Db\Pd\{ D,\Db \} \{ D,\Db \}(\P + \Pd)  \Big\vert_{\th\th\tb\tb}  \nn\\ 
\space \nn \\
&=&
\frac{1}{4\phi^2} \pa_\mu (\pa \phi)^2 \pa^\mu (\pa \phi)^2 
-\frac{1}{2\phi^2} \phi^{,\mu} \pa_{\mu} (\pa \phi)^2 \square \phi
+ \frac{1}{4\phi^3} (\pa \phi)^4 \square \phi
\nn \\
&-& \frac{1}{2\phi^2} \pa_\mu (\pa \phi)^2 \pa^\mu (\pa \chi)^2 
+ \frac{1}{2\phi^2} \pa_\mu (\pa \phi \cdot \pa \chi) \pa^\mu (\pa \phi \cdot \pa \chi) 
+ \frac{1}{4\phi^2} \pa_\mu (\pa \chi)^2 \pa^\mu (\pa \chi)^2
\nn \\
&+& \frac{1}{2\phi^2} \phi^{,\mu} \pa_{\mu} (\pa \chi)^2 \square \phi 
+ \frac{1}{\phi^2} \phi^{,\mu} \pa_{\mu} (\pa \phi \cdot \pa \chi) \square \chi
\nn \\
&+& \frac{1}{4 \phi^3} (\pa \chi)^4 \square \phi 
- \frac{1}{2\phi^3}(\pa \phi)^2 (\pa \chi)^2 \square \phi 
+ \frac{1}{\phi^3}(\pa \phi \cdot \pa \chi)^2 \square \phi
\nn \\
&+& \bigg[
- \frac{1}{\phi^2}  \phi_{,\mu \nu} \phi^{, \mu \nu} - \frac{3}{\phi^2} \chi_{,\mu \nu} \chi^{, \mu \nu} + \frac{1}{\phi^3} \phi^{,\mu}\pa_\mu \big(  (\pa \phi)^2  -  (\pa \chi)^2 \big) + \frac{2}{\phi^3} \chi_{,\mu} \chi^{,\mu \nu} \phi_{,\nu}
\nn \\
&+& \frac{1}{\phi^2} \phi^{,\mu} \square \big( \phi_{,\mu} + \chi_{,\mu}\big)- \frac{1}{\phi^3} \big( (\pa \phi)^2 \square \phi + (\pa \chi)^2 \square \phi \big) 
\bigg] F F^*
\nonumber
\end{eqnarray}
\begin{eqnarray}
~~~~~~~~~~~~~~&+& \bigg[- \frac{1}{\phi^2} \big( \phi_{,\mu} \phi^{,\mu \nu}  + \chi_{,\mu} \chi^{,\mu \nu}  \big)+\frac{1}{\phi^3}\phi^{,\mu}(\pa \chi)^2 
- \frac{i}{\phi^3} \phi^{,\mu}(\pa \phi \cdot \pa \chi)
+\frac{1}{2\phi^2} \phi^{,\mu} \square ( \phi + \chi ) 
\nn
\\
~~~~~~~~~~~~~~&-& \frac{1}{4\phi^2} \big( (\pa \phi)^2 - (\pa \chi)^2 + 2i \pa \phi \cdot \pa \chi \big)^{,\mu}
+ \frac{1}{2\phi^2} \chi^{,\mu \nu} \big(  \chi_{,\nu} - i \phi_{,\nu} \big)
\label{32} \\
~~~~~~~~~~~~~~&+& \frac{1}{2\phi^2} \phi_{,\nu}\big( \phi^{,\mu \nu} - i \chi^{,\mu \nu}\big)
\bigg]  F^* \partial_{\mu}F 
\nn \\
~~~~~~~~~~~~~~&+& \bigg[
- \frac{1}{\phi^2} \big( \phi_{,\mu} \phi^{,\mu \nu}  + \chi_{,\mu} \chi^{,\mu \nu}  \big)
+\frac{1}{\phi^3}\phi^{,\mu}(\pa \chi)^2 
- \frac{i}{\phi^3} \phi^{,\mu}(\pa \phi \cdot \pa \chi)
+\frac{1}{2\phi^2} \phi^{,\mu} \square ( \phi + \chi ) 
\nn \\
~~~~~~~~~~~~~~&-&\frac{1}{4\phi^2} \big( (\pa \phi)^2 - (\pa \chi)^2 + 2i \pa \phi \cdot \pa \chi \big)^{,\mu}
+ \frac{1}{2\phi^2} \chi^{,\mu \nu} \big(  \chi_{,\nu} + i \phi_{,\nu} \big)
\nn \\ 
~~~~~~~~~~~~~~&+&\frac{1}{2\phi^2} \phi_{,\nu}\big( \phi^{,\mu \nu} + i \chi^{,\mu \nu}\big)
\bigg]F \partial_{\mu}F^*
\nn \\
~~~~~~~~~~~~~~&+&\frac{1}{\phi^2}\big( (\pa \phi)^2  + (\pa \chi)^2 \big) \pa F^* \cdot \pa F
+ \frac{1}{\phi^2} \phi^{,\mu} \phi^{,\nu} \partial_{\mu}F \partial_{\nu} F^* 
+ \frac{i}{2\phi^2}\big(
\chi^{,\mu } \phi^{,\nu}  
- \chi^{,\nu } \phi^{,\mu}  
\big)\partial_{\mu}F^* \partial_{\nu}F
\nn \\
~~~~~~~~~~~~~~&+& \frac{6}{\phi^2} F^* F \pa F^* \cdot \pa F
+ \frac{1}{\phi^3} F (F^*)^2 (\pa F \cdot \pa \phi)
+ \frac{1}{\phi^3} F^* F^2 (\pa F^* \cdot \pa \phi)
+ \frac{1}{\phi^3} \square \phi(F F^*)^2 . \nonumber
\nonumber
\end{eqnarray}
Expressed in terms of the complex scalar field $A=\frac{1}{\sqrt{2}}(\phi+i\chi)$, this can be re-written as
\bea
{\bar{\mathcal{L}}}_{4}^{\mathrm{SUSY}}&=&\frac{1}{(A+A^{*})^{2}}\bigg[2\partial_{\mu}(\partial A)^{2} \partial^{\mu}(\partial{A}^{*})^{2} 
-4|\partial A |^{2} |\partial F|^{2}  +12|F|^{2}|\partial F|^{2}
\nn \\
&+&
\bigg( \big(
2 \partial^{\mu}F \Box A  + \pt^{\mu} (\pt A \cdot \pt F) 
\big) F^* +\big(2 \partial^{\mu}F^* \Box A^*  + \pt^{\mu} (\pt A^* \cdot \pt F^*) 
\big) F \bigg)\partial_{\mu}(A + A^*)
\nn \\
\nn \\
&+&\big(-6 \partial_{\mu} \partial_{\nu}A\partial^{\mu}\partial^{\nu}A^{*} 
+  
\partial^{\mu}\partial^{\nu}A \partial_{\mu}\partial_{\nu} A 
+\partial^{\mu}\partial^{\nu}A^{*} \partial_{\mu}\partial_{\nu} A^{*} 
\big) |F|^{2} 
\nonumber \\
\nn \\ 
&-& 
 \pt^{\mu }((\pt A)^2 ) \partial_{\mu}F F^* 
-  \pt^{\mu }((\pt A^*)^2 ) \partial_{\mu}F^* F 
\nn \\
\nn \\
&-&
3\big( \partial_{\nu}A\partial^{\nu}\partial^{\mu}A^{*}(\partial_{\mu}F)F^{*} 
+\partial_{\nu}A^{*}\partial^{\nu}\partial^{\mu}A F\partial_{\mu}F^{*}\big)
\nonumber \\
\nn \\
&-&
\partial_{\mu}\partial_{\nu}A \partial^{\nu}A \partial^{\mu}FF^* 
-\partial_{\mu} \partial_{\nu}A^{*} \partial^{\nu}A^{*} \partial^{\mu}F^*F 
\nn \\
\nn \\
&+& \bigg(\big(
\partial_{\nu}F^* \partial^ {\nu}\partial^{\mu}A F + \partial_{\nu}F^* \partial^{\nu}A \partial^{\mu}F
\big) 
+ \big(
\partial_{\nu}F \partial^ {\nu}\partial^{\mu}A^{*} F^{*} + \partial_{\nu}F \partial^{\nu}A^{*} \partial^{\mu}F^{*}
\big)
\nn \\
&-&  
\pt^{\mu}(\pt A)^2 \Box A^*
- \pt^{\mu}(\pt A^*)^2 \Box A \Bigg)\partial_{\mu}( A + A^*)
\bigg]
\nn \\
&+&
\frac{2}{(A + A^*)^{3}} \bigg[
\bigg( \big(
 \pt^\mu (\pt A)^2 +  \pt^\mu(\pt A^*)^2 
- 2 \partial^{\mu}F F^* - 2 \partial^{\mu}F^{*} F\big)|F|^{2}
\nn \\
&-&
( \partial^{\nu} \partial^{ \mu}A F + \partial^{\nu}A \partial^{\mu} F) F^* 
+( \partial^{\nu} \partial^{ \mu}A^* F^* + \partial^{\nu}A^* \partial^{\mu} F^*) F 
\bigg) \partial_{\nu}(A - A^*)\partial_{\mu}(A + A^*)
\nn \\
&+&\bigg((\partial A)^{2}(\partial A^{*})^{2}-2(\partial A \cdot \partial A^{*})|F|^{2}+ |F|^{4}\bigg)\Box (A+A^{*}) \bigg]
\label{bw7}
\eea

\subsection{${\bar{\cal{L}}}_{1}$}

In the preceding subsections, we have presented both the superfield and component field expressions--ignoring the fermion--for the supersymmetrization of ${\bar{{\cal L}}}_2$, ${\bar{{\cal L}}}_3$ and ${\bar{{\cal L}}}_4$ in \eqref{22}. However, there is also a pure potential term ${\bar{{\cal L}}}_1$ in \eqref{22}. How does one supersymmetrize it? As is well known, this is accomplished by adding a superpotential $W$ to the superfield Lagrangian. $W$ is a holomorphic function of the chiral superfield $\Phi$ introduced above. It follows that its supersymmetric Lagrangian, which we denote by ${\bar{{\cal L}}}^{SUSY}_1$, is given by
\be
{\bar{{\cal L}}}^{SUSY}_1 = W|_{\theta\theta} + W^{\dagger}|_{{\bar{\theta}}{\bar{\theta}}}= \frac{\partial W}{\partial A}F + \frac{\partial W^*}{\partial A^*}F^*\ .
\label{33}
\ee
In the component field expression on the right-hand side, $W=W(A)$ with $A=\frac{1}{\sqrt{2}}(\phi+i\chi)$, as defined above. Note that since we are taking ${\bar{{\cal L}}}^{SUSY}_1$ and scalar field $A$ to be dimensionless, and since $F$ (recall, really $\frac{F}{\cal{M}}$) has dimension 0, it follows that the superpotential $W$ (really $\frac{W}{{\cal{M}}^{3}})$ must also have mass dimension 0.

Having introduced the superpotential term, we can now write the entire\footnote{Recall that we are ignoring, for simplicity, the ${\cal{L}}_{5}$ Galileon.} supersymmetric Lagrangian for the worldvolume action of a 3-brane in $AdS_{5}$ 5-space with an $M_{4}$ foliation. It is given by
\be
{\cal{L}}^{SUSY}=\Sigma_{i=1}^{4} {\bar{c}}_{i}{\bar{\cal{L}}}^{SUSY}_{i} \ ,
\label{34}
\ee
 where ${\bar{\cal{L}}}^{SUSY}_{i}$, $i=1,2,3,4$ are given by expressions \eqref{33}, \eqref{25}, \eqref{28} and \eqref{32} respectively. The constant coefficients ${\bar{c}}_{i}$ are all real, each has dimension 4 but are otherwise arbitrary.

\subsection{The $F$-Field Terms in the Lagrangian}

In this section, we will isolate and discuss only those terms in ${\cal{L}}^{SUSY}$ that contain at least one $F$-field. This will be denoted by ${\cal{L}}^{SUSY}_{F} \subset {\cal{L}}^{SUSY}$,
and is given by
\bea
{\cal{L}}^{SUSY}_{F} &=&
\bar{c}_1\frac{\partial W}{\partial A}F + \bar{c}_1\frac{\partial W^*}{\partial A^*}F^* 
+ \bar{c}_2 \frac{1}{\phi^{4}}F^{*}F
\nn \\
&&+ \bar{c}_3 \big( -\frac{1}{\phi^3} \square \phi + \frac{3}{2\phi^4}(\pa \phi)^2  + \frac{3}{2\phi^4}(\pa \chi)^2\big)F^* F
+ \bar{c}_3 \frac{i}{\phi^3} \chi_{, \mu} (F^* \partial^{\mu}F - F \partial^{\mu}F^*)
\nn \\
&&+\bar{c}_4 \bigg[
- \frac{1}{\phi^2}  \phi_{,\mu \nu} \phi^{, \mu \nu} - \frac{3}{\phi^2} \chi_{,\mu \nu} \chi^{, \mu \nu} + \frac{1}{\phi^3} \phi^{,\mu}\pa_\mu \big(  (\pa \phi)^2  -  (\pa \chi)^2 \big) + \frac{2}{\phi^3} \chi_{,\mu} \chi^{,\mu \nu} \phi_{,\nu}
\nn \\
&& 
+\frac{1}{\phi^2} \phi^{,\mu} \square \big( \phi_{,\mu} + \chi_{,\mu}\big)- \frac{1}{\phi^3} \big( (\pa \phi)^2 \square \phi + (\pa \chi)^2 \square \phi \big) 
\bigg] F F^*
\nn \\
\nn \\
&&+\bar{c}_4 \bigg[
- \frac{1}{\phi^2} \big( \phi_{,\mu} \phi^{,\mu \nu}  + \chi_{,\mu} \chi^{,\mu \nu}  \big)
+\frac{1}{\phi^3}\phi^{,\mu}(\pa \chi)^2 
- \frac{i}{\phi^3} \phi^{,\mu}(\pa \phi \cdot \pa \chi)
+\frac{1}{2\phi^2} \phi^{,\mu} \square ( \phi + \chi ) 
\nn \\
\nn \\
&& 
- \frac{1}{4\phi^2} \big( (\pa \phi)^2 - (\pa \chi)^2 + 2i \pa \phi \cdot \pa \chi \big)^{,\mu}
+ \frac{1}{2\phi^2} \chi^{,\mu \nu} \big(  \chi_{,\nu} - i \phi_{,\nu} \big)
\nn \\
\nn \\
\nonumber
\eea
\bea
&& 
\qquad\qquad +\frac{1}{2\phi^2} \phi_{,\nu}\big( \phi^{,\mu \nu} - i \chi^{,\mu \nu}\big)
\bigg]  F^* \partial_{\mu}F
\nonumber 
\label{bird2} 
\\
&&\qquad\qquad+\bar{c}_4 \bigg[
- \frac{1}{\phi^2} \big( \phi_{,\mu} \phi^{,\mu \nu}  + \chi_{,\mu} \chi^{,\mu \nu}  \big)
+\frac{1}{\phi^3}\phi^{,\mu}(\pa \chi)^2 
+ \frac{i}{\phi^3} \phi^{,\mu}(\pa \phi \cdot \pa \chi)
+\frac{1}{2\phi^2} \phi^{,\mu} \square ( \phi + \chi ) 
\nn \\
\nn \\
&& 
\qquad\qquad- \frac{1}{4\phi^2} \big( (\pa \phi)^2 - (\pa \chi)^2 - 2i \pa \phi \cdot \pa \chi \big)^{,\mu}
+ \frac{1}{2\phi^2} \chi^{,\mu \nu} \big(  \chi_{,\nu} +i \phi_{,\nu} \big)
\nn \\
\nn \\
&& 
\qquad\qquad+\frac{1}{2\phi^2} \phi_{,\nu}\big( \phi^{,\mu \nu} + i \chi^{,\mu \nu}\big)
\bigg]F \partial_{\mu}F^*
 \\
 \label{bird2}
&& \qquad\qquad+\bar{c}_4\frac{1}{\phi^2}\big( (\pa \phi)^2  + (\pa \chi)^2 \big) \pa F^* \cdot \pa F
+\bar{c}_4\frac{1}{\phi^2} \phi^{,\mu} \phi^{,\nu} \partial_{\mu}F \partial_{\nu} F^* 
+\bar{c}_4\frac{i}{2\phi^2}\big(
\chi^{,\mu } \phi^{,\nu}  
- \chi^{,\nu } \phi^{,\mu}  
\big)\partial_{\mu}F^* \partial_{\nu}F
\nn \\
&&\qquad\qquad +\bar{c}_4 \frac{6}{\phi^2} F^* F \pa F^* \cdot \pa F
+\bar{c}_4 \frac{1}{\phi^3} F (F^*)^2 (\pa F \cdot \pa \phi)
+ \bar{c}_4\frac{1}{\phi^3} F^* F^2 (\pa F^* \cdot \pa \phi)
\nn \\
\nn \\
&&\qquad\qquad+ \bar{c}_4\frac{1}{\phi^3} \square \phi(F F^*)^2 \nonumber
\nn 
\eea

Clearly, the $F$-field is no longer a simple auxiliary field. There are two reasons for this. The first is that, in addition to terms proportional to $F$, $F^*$ and $F^*F$, there are also terms of order $(F^*F)^{2}$. Secondly, there are terms with both a single derivative $\partial F$ or $\partial F^*$, as well as terms containing two derivatives such as $\partial F^{*}\partial F$. Assuming, for a moment, that there are no terms with derivatives acting on $F$ or $F^*$, such Lagrangians would lead to a cubic equation for $F$. This would have three inequivalent solutions and lead, for example, to three different expressions when $F$ is inserted back into the Lagrangian. For example, these Lagrangians would have different potential energy functions. This has previously been explored in several contexts in \cite{Koehn:2012ar, Ciupke:2015msa}. Perhaps more intriguing is the second case when the Lagrangian contains $F$ terms with one or more derivatives acting on them. This would imply that the $F$-field is dynamical and no longer a true auxiliary field. There is nothing wrong with this from the point of view of supersymmetry representations--an irreducible supermultiplet containing two dynamical complex scalars $A$ and $F$, each paired with a Weyl spinor, does exist. The dynamics of such theories, to our knowledge, has only been discussed in the trivial case where the superpotential $W$ is zero and, hence, one can take $F=0$ as the solution \cite{Koehn:2013upa}. In the present paper, however, we will deal directly with the issue of derivatives on the field $F$, and its elimination from the Lagrangian, in the non-trivial case where the superpotential does not vanish. We do this as follows.

First, recall that we have obtained the conformal Galileons in \eqref{18} by doing an expansion of the DBI conformal Galileons \eqref{14} in powers of $(\frac{\partial}{\cal M})^{2}$--where we have momentarily restored the mass ${\cal M}=1/{\cal{R}}$. Since terms in this expansion with $(\frac{\partial}{\cal M})^{2p}$ where $p > 4$ are a total divergence, it is not strictly necessary to assume that $(\frac{\partial}{\cal M})^{2} \ll 1$ . Be that as it may, since we would like to ignore ${\bar{\cal{L}}}_{5}$ in our calculations and, furthermore, simplify the discussion of the $F$-field, we will henceforth assume
\be 
(\frac{\partial}{\cal M})^{2} \ll 1 \ .
\label{r1}
\ee
Additionally, note that the supersymmetric $\bar{\mathcal{L}}_{3}^{\mathrm{SUSY}}$ and $\bar{\mathcal{L}}_{4}^{\mathrm{SUSY}}$ Lagrangians in \eqref{28} and \eqref{32} respectively contain term involving the field $F$ and its derivatives. For example, consider ${\bar{\cal{L}}}_{3}$. By definition, this Lagrangian contains pure scalar terms with four powers of derivatives, such as $\frac{1}{2\phi^3} (\pa \phi)^2 \square \phi$. However, the same Lagrangian contains terms, for example $\frac{3}{2\phi^4}(\pa \phi)^2F^* F$, which have two powers of derivatives acting on scalars multiplied by $F^{*}F$. 
It is natural, and greatly simplifies our analysis, if we demand that all terms in each of the $\bar{\mathcal{L}}_{i}^{\mathrm{SUSY}}$ Lagrangians be of the same order of magnitude. Since the terms involving derivatives satisfy \eqref{r1}, it follows that one must also choose
\be
\bigg|\frac{F}{\cal{M}}\bigg|^{2} \ll 1 \ ,
\label{ssA}
\ee
which we assume henceforth.

With this in mind, it is reasonable to solve the equation for $F$ as an expansion in $(\frac{\partial}{\cal M})^{2}$ as well. To zeroth order in this expansion, the relevant part of Lagrangian \eqref{bird2} becomes
\be
{\cal{L}}^{SUSY(0)}_{F} =
\bar{c}_1\frac{\partial W}{\partial A}F^{(0)} + \bar{c}_1\frac{\partial W^*}{\partial A^*}F^{0*} 
+ \bar{c}_2 \frac{1}{\phi^{4}}F^{(0)*}F^{(0)} \ .
\label{36}
\ee
The equation of motion for $F^{(0)*}$ then implies that
\be
F^{(0)}=-\frac{{\bar{c}}_1}{{\bar{c}}_2} \phi^4 \frac{\partial W^*}{\partial A^*} \ .
\label{37}
\ee
Putting this expression back into \eqref{36} yields
\be
{\cal{L}}^{SUSY(0)}_{F}= -\frac{{\bar{c}}_{1}^{2}}{\bar{c}_{2}}{\phi}^{4} \bigg |  \frac{\partial W}{\partial A} \bigg |^{2} \ .
\label{38}
\ee
Since $W$ is a holomorphic function of the complex scalar field $A=\frac{1}{\sqrt{2}}(\phi+i\chi)$, it follows that the above expression is simply minus the potential energy 
\be
V=\frac{{\bar{c}}_{1}^{2}}{\bar{c}_{2}}{\phi}^{4} \bigg |  \frac{\partial W}{\partial A} \bigg |^{2} \ .
\label{39}
\ee
One now must choose the form of $W$ so that, when one sets $\chi=0$, the potential becomes $V=\frac{{\bar{c}}_{1}}{4\phi^{4}}$, as required by ${\bar{{\cal L}}}^{SUSY}_1$ in \eqref{22}.
This is easily satisfied if one chooses 
\be
W=\frac{1}{2}\sqrt{ \frac{{\bar{c}}_{2}}{ {\bar{c}}_{1}} } \frac{1}{12A^{3}} \ .
\label{40}
\ee
It follows from this and \eqref{39} that
\be
V=\frac{{\bar{c}}_{1}}{4} \frac{\phi^{4}}{(\phi^{2}+\chi^{2})^{4}} \ .
\label{41}
\ee
Hence, for $\chi=0$ this reproduces the potential in ${\bar{{\cal L}}}^{SUSY}_1$, as required. It is also of interest to note that for this choice of $W$
\be
F^{(0)}=-\frac{1}{2}\sqrt{ \frac{{\bar{c}}_1}{{\bar{c}}_2} } \frac{\phi^{4}}{(\phi-i\chi)^{4}} \ .
\label{42}
\ee
It follows that when $\chi=0$ the field $F^{(0)}$ is a constant. 

Henceforth, in this paper, we would like to restrict ourselves to the subset of solution space where 
\be
\chi=\partial_{\mu} \chi =0 \ .
\label{43}
\ee
First, of course, one must show that such solutions are possible. We begin by considering the form of potential \eqref{41}. For positive values of coefficient ${\bar{c}}_{1}$, the potential is everywhere positive and, for any fixed value of $\phi$, it is minimized as $\chi \rightarrow \pm \infty$. Furthermore, at $\chi=0$, the mass $m_{\chi}^{2} =\frac{\partial^{2}V}{{\partial \chi}^{2}}|_{\chi=0}=-\frac{2{\bar{c}}_1}{\phi^{6}} < 0$ for any value of $\phi$. Hence, the solution $\chi=0$ to the equations of motion would be fine-tuned and highly unstable. This unsatisfactory situation can easily be corrected by simply imposing the condition that
\be 
{\bar{c}}_1 < 0 \ ,
\label{44}
\ee
which we do henceforth. Potential \eqref{41} then becomes everywhere negative--which, for example, is required in bouncing universe cosmological scenarios. It follows that for any value of $\phi$ the potential grows larger as $\chi \rightarrow \pm \infty$ and 
\be
m_{\chi}^{2}=\frac{\partial^{2}V}{{\partial \chi}^{2}}|_{\chi=0}=\frac{2|{\bar{c}}_1|}{\phi^{6}} > 0 \ .
\label{45}
\ee
The potential energy $V$ in a range of $\phi$-$\chi$ space is shown in Figure{~1}.
\begin{figure}[h]
\begin{center}
\includegraphics[scale=0.3]{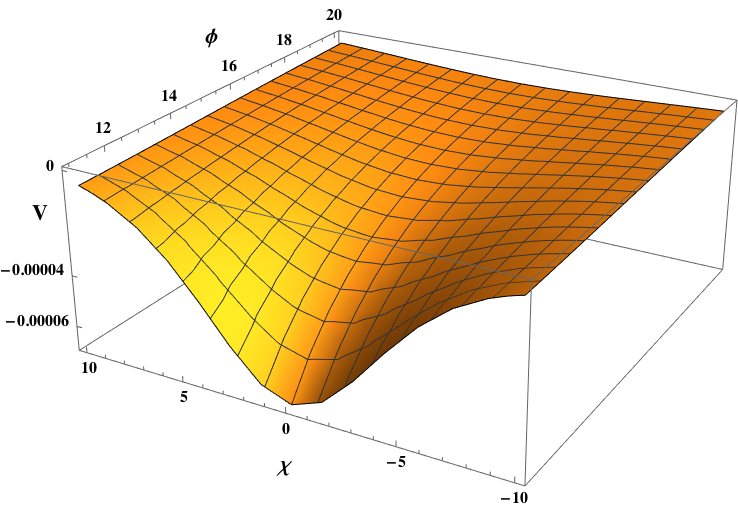}
\caption{The potential energy $V$ plotted over a region of $\phi$-$\chi$ space.}
\label{1}
\end{center}
\end{figure}
Therefore, if the coefficients are chosen so that the kinetic energy term for $\chi$ is non-ghost like, which we will impose below, then solutions where $\chi=\partial_{\mu}\chi=0$ are indeed possible. Note from \eqref{42} that, in this case,
\be
F^{(0)}=\frac{i}{2}\sqrt{ \frac{|{\bar{c}}_1|}{{\bar{c}}_2} } \ .
\label{46}
\ee
Note that in order for $F^{(0)}$ to satisfy \eqref{ssA}, then
\be
F^{*(0)}F^{(0)} \ll 1 \quad \Longrightarrow \quad {\bar{c}}_{2} \gg |{\bar{c}}_{1}| \ .
\label{r4}
\ee
Taking \eqref{44}, and putting the constant $F^{(0)}$-field \eqref{46} back into Lagrangian \eqref{bird2}, we find that ${\cal{L}}^{SUSY}_{F}$ simplifies to
\bea
{\cal{L}}^{SUSY}_{F} &=& -\frac{{\bar{c}}_{1}}{4\phi^{4}}
\nn \\
&+& \bar{c}_3 \big( - \frac{3}{2\phi^4}(\pa \phi)^2  + \frac{3}{2\phi^4}(\pa \chi)^2\big)F^{(0)*} F^{(0)}
\nn \\
&+&\bar{c}_4 \big(
+\frac{1}{\phi^2} \phi^{,\mu} \square \big( \phi_{,\mu} + \chi_{,\mu}\big) -\frac{3}{\phi^{4}}(\partial \phi)^{4}
\big) F^{(0)*} F^{(0)} 
\nn \\
&+& 3\bar{c}_4\frac{1}{\phi^4} (\partial\phi)^{2}(F^{(0)*} F^{(0)})^2 \ .
\label{47}
\eea
In deriving this expression, we have dropped all terms containing at least one power of $\partial_{\mu}F^{(0)}$, have kept the terms containing $\chi$ only where a) it would could contribute to the lowest order $\chi$ kinetic energy term or b) is linear in $\chi$ and, hence, its variation does not vanish in the $\chi$ equation of motion when one sets $\chi=\partial_{\mu}\chi=0$. Furthermore, we have simplified the remaining expressions using integrating by parts, which we can do since $\partial_{\mu} F^{(0)}=0$. All other terms containing $\chi$ in ${\cal{L}}^{SUSY}_{F}$ are quadratic in $\chi$ and would vanish in both the $\phi$ and $\chi$ equations of motion and, hence, can be dropped from the Lagrangian.

The constraint \eqref{44} ensuring that the potential energy, for fixed $\phi$, is minimized at $\chi=0$ is not the only constraint that one might put on the coefficients ${\bar{c}}_{i}$.
Depending on the physical problem being analyzed, the coefficients of the two-derivative kinetic energy terms for both $\phi$ and $\chi$, 
which generically depend on ${\bar{c}}_{i}$ and $F^{(0)}$, must be appropriately chosen. 
Adding the first two terms of ${\bar{\cal{L}}}_{2}^{\rm SUSY}$ in \eqref{25} to the $\frac{1}{\phi^{4}}(\partial \phi)^{2}$ and 
$\frac{1}{\phi^{4}}(\partial \chi)^{2}$ terms in ${\cal{L}}^{SUSY}_{F}$ in \eqref{47} yields
\bea
&-& \frac{1}{2\phi^{4}}(\partial \phi)^{2} \big( {\bar{c}}_{2}  + 3{\bar{c}}_{3}F^{(0)*}F^{(0)} -6{\bar{c}}_{4}\big(F^{(0)*}F^{(0)}\big)^{2}  \big)
\nn \\
&-& \frac{1}{2\phi^{4}}(\partial \chi)^{2} \big({\bar{c}}_{2}  - 3{\bar{c}}_{3}F^{(0)*}F^{(0)}  \big) \ .
\label{48}
\eea
As discussed above, to obtain solutions of the $\chi$ equation of motion for which $\chi=\partial_{\mu}\chi=0$ requires that the $\chi$ kinetic energy be ghost free. It follows from the second term in \eqref{48} that one must therefore impose 
\be
{\bar{c}}_{2}  - 3{\bar{c}}_{3}F^{(0)*}F^{(0)} > 0 \ .
\label{r3}
\ee
On the other hand, the sign and magnitude of the coefficient of the $\phi$ kinetic energy term depends on the type of physics one is interested in. For example, if one wants the $\phi$ field to develop a ``ghost condensate''--one way in which the null energy condition (NEC) can be violated--then it follows from the first term in \eqref{48} that the coefficients ${\bar{c}}_{i}$ should be chosen so that 
\be
{\bar{c}}_{2}  + 3{\bar{c}}_{3}F^{(0)*}F^{(0)} -6{\bar{c}}_{4}\big(F^{(0)*}F^{(0)}\big)^{2} < 0 \ .
\label{49}
\ee
However, it is well-known \cite{ Hinterbichler:2012yn, Khoury:2011da} that Galileon Lagrangians can, for appropriate choices of coefficients, 
violate the NEC without developing a ghost condensate. In such cases, one can choose the coefficient of the $\phi$ kinetic energy to be positive.

With this in mind, we now calculate the field $F$ to first order in the derivative expansion. Denoting
\be
F^{((0)+(1))}=F^{(0)}+F^{(1)} \ ,
\label{r6}
\ee
where $F^{(0)}$ is computed from the zeroth order Lagrangian \eqref{36},
the relevant part of ${\cal{L}}^{SUSY}_{F}$ then becomes
\bea
{\cal{L}}^{SUSY ( (0)+(1) )}_{F} &=&
\bar{c}_1\frac{\partial W}{\partial A}F^{((0)+(1))} + \bar{c}_1\frac{\partial W^*}{\partial A^*}F^{*((0)+(1))}
+ \bar{c}_2 \frac{1}{\phi^{4}}F^{*((0)+(1))}F^{((0)+(1))}
\nn \\
&+& \bar{c}_3 \big( -\frac{1}{\phi^3} \square \phi + \frac{3}{2\phi^4}(\pa \phi)^2  + \frac{3}{2\phi^4}(\pa \chi)^2\big)F^{*((0)+(1))}F^{((0)+(1))} \nn \\
&+& \bar{c}_3 \frac{i}{\phi^3} \chi_{, \mu} (F^{*((0)+(1))} \partial^{\mu}F^{((0)+(1))} - F^{((0)+(1))} \partial^{\mu}F^{*((0)+(1))}) \ .
\label{r7}
\eea
Now insert \eqref{r6} into \eqref{r7}.  Recalling that we are always choosing constraints \eqref{44} and \eqref{r3}  so that $\chi=\partial_{\mu} \chi =0$, it follows that $F^{(0)}$ is given by the constant \eqref{46} and, hence, $\partial_{\mu}F^{(0)}=0$. 
Dropping terms proportional to $F^{*(1)}F^{(1)}$ (which are of second order in the $F$ expansion) and integrating the last term by parts using the fact that $\partial_{\mu}F^{(0)}=0$, we find that 
\bea
F^{(1)} &=&
- \frac{\bar{c}_3}{\bar{c}_2} \phi^4 F^{(0)}  
\big( -\frac{1}{\phi^3} \square \phi + \frac{3}{2\phi^4}(\pa \phi)^2  + \frac{3}{2\phi^4}(\pa \chi)^2 +i( \frac{1}{\phi^{3}} \square \chi -3\frac{1}{\phi^{4}}\partial^{\mu}\phi \partial_{\mu}\chi) \big) 
\nn \\
&=&-{\frac{i}{2}}
\frac{{\bar{c}}_3}{{\bar{c}}_2}
\sqrt{\frac{|{\bar{c}}_{1}|}{{\bar{c_{2}}}}} 
\big( -\phi \square \phi + \frac{3}{2}(\pa \phi)^2  + \frac{3}{2}(\pa \chi)^2+i( \phi \square \chi -3\partial^{\mu}\phi \partial_{\mu}\chi)\big) \ .
\label{r8}
\eea
It follows that 
\be
F^{((0)+(1))}= F^{(0)} \big(1 - \frac{{\bar{c}}_3}{{\bar{c}}_2}   
\big( -\phi \square \phi + \frac{3}{2}(\pa \phi)^2  + \frac{3}{2}(\pa \chi)^2 +i( \phi \square \chi -3\partial^{\mu}\phi \partial_{\mu}\chi) \big)  \big)
\label{r9}
\ee
Clearly, the $F^{(1)}$ term in this expansion is small compared to the $F^{(0)}$ term, since we are working in the limit where $(\partial)^{2} \ll 1 $. One can insert \eqref{r9} back into the $F$-term Lagrangian ${\cal{L}}^{SUSY}_{F} $ in \eqref{bird2}, as we did in the zeroth order case. However, as far as the analysis in this paper is concerned, there is nothing to be gained from doing this--simply more yet higher derivative terms. Hence, we will not do that here--contenting ourselves with the zeroth order $F$-term Lagrangian given in \eqref{47}. 

It is clear, however, that one can consistently do  a higher order expansion to completely determine the perturbative solution for the $F$-field, up to and including terms arising in ${\bar{\cal{L}}}_{5}$. Here, we demonstrate this by computing the next order, $F^{(2)}$, in the $(\partial)^{2}$ expansion of $F$. Denoting
\be
F^{((0)+(1)+(2))}=F^{(0)}+F^{(1)}+F^{(2)} \ ,
\label{ho1}
\ee
where $F^{(0)}$ and $F^{(1)}$ are determined from \eqref{36} and \eqref{r7} respectively, the relevant part of ${\cal{L}}^{SUSY}_{F}$ is given by the entire expression \eqref{bird2}
with $F$ replaced by $F^{((0)+(1)+(2))}$. Differentiating this with respect to $F^{*}$, where in terms involving $\partial_{\mu}F^{*}$ we use integration by parts to remove a derivative,
one arrives at the equation of motion for $F^{((0)+(1)+(2))}$. This is solved for $F^{(2)}$ as follows. First recall that $F^{(0)}$ is given by \eqref{37}. As discussed above, we will always choose constraints \eqref{44} and \eqref{r3}  so that $\chi=\partial_{\mu} \chi =0$. Hence, it follows that $F^{(0)}$ is given by the constant \eqref{46} and, therefore, that $\partial_{\mu}F^{(0)}=0$. Second, recall that the expression for $F^{(1)}$ is presented in \eqref{r8}. Insert  \eqref{37}, \eqref{46} and \eqref{r8} into the equation for $F$.  Third, drop all terms in the 
$F$ equation involving $\partial ^{n}, n>4$. Finally, recall that we have imposed \eqref{ssA} on the magnitude of the $F$ field. Using this, it follows that terms of the form, say, $\Box \phi ~|F^{(0)}|^{2}F^{(0)}$ have the appropriate dimension whereas terms like $\Box \phi ~|F^{(0)}|^{2}F^{(1)}$ do not, even though this last term is proportional to $\partial^{4}$. Putting everything together, we find that
\bea
F^{(2)} &=& F^{(0)-1}(F^{(1)})^{2} - \frac{2{\bar{c}}_{3}}{{\bar{c}}_{2}} \phi \partial_{\mu}\chi \partial^{\mu} F^{(1)} \nn \\
&+& \frac{{\bar{c}}_{4}}{{\bar{c}}_{2}} \bigg(\phi^2  \phi_{,\mu \nu} \phi^{, \mu \nu}  +3 \phi^2 \chi_{,\mu \nu} \chi^{, \mu \nu} - \phi \phi^{,\mu}\pa_\mu \big(  (\pa \phi)^2  -  (\pa \chi)^2 \big) -2 \phi \chi_{,\mu} \chi^{,\mu \nu} \phi_{,\nu}
\nn \\
&-&\phi^2 \phi^{,\mu} \square \big( \phi_{,\mu} + \chi_{,\mu}\big) + \phi \big( (\pa \phi)^2 + (\pa \chi)^2 \big) \square \phi   \bigg) F^{(0)} \nn \\
&-&\frac{{\bar{c}}_{4}}{{\bar{c}}_{2}} \big( \phi \Box \phi +3(\partial \phi)^{2} \big) |F^{(0)}|^{2} F^{(0)} \ .
\label{ho2}
\eea
There is one important caveat in arriving at \eqref{ho2}.  The expression for $F^{(2)}$ actually, in addition to the above, contains a term which is $\phi^{4}$ times a total derivative. When inserted back into the Lagrangian, this would be higher order in all terms proportional to ${\bar{c}}_{3}$ and ${\bar{c}}_{4}$ and, hence, can be ignored. However, it would have to be included in the terms proportional to ${\bar{c}}_{1}$ and ${\bar{c}}_{2}$. However, inserting it into these terms, integrating by parts and recalling that we will always solve the equations of motion so that $\chi=\partial_{\mu} \chi =0$, it follows that these contributions exactly vanish. Hence, we have dropped this total derivative term from the expression \eqref{ho2} for $F^{(2)}$.

%%%%%%%%%%%%%%%%%%%%%%%%%%%

\section{Extension of Conformal Galileons to $N=1$ Supergravity}

In the previous section, we extended the flat space conformal Galileons of a single real scalar field to $N=1$ supersymmetry. We now further generalize these results by extending them to curved $N=1 $ superspace and, hence, to $N=1$ supergravity. This is accomplished by using, and then greatly expanding upon, results on higher-derivative supergravitation first presented in \cite{Koehn:2012ar, Baumann:2011nm, Farakos:2012je, Farakos:2012qu}. Throughout this section we use results from, and follow the notation of, the book ``Supersymmetry and Supergravity'' by Bagger and Wess \cite{Wess}.\footnote{However, our index labelling convention differs from \cite{Wess} in the following way. Tangent space bosonic and spinor indices are chosen from the start of the Latin and Greek alphabet respectively--e.g. $a$, $b$ and $\alpha$, $\dot{\alpha}$ .
Spacetime \emph{bosonic} indices are taken from the middle of the \emph{Greek} alphabet--e.g. $\mu$, $\nu$. We will not deal with spacetime spinor indices in this section.
}

We begin with a purely chiral superfield $\Phi (x, \theta, \bar{\theta})$ in flat superspace. By definition, this satisfies the constraint ${\bar{D}}_{\dot{\alpha}}\Phi=0$, where ${\bar{D}}$ is the flat superspace differential operator. Note that one can generically construct a Lagrangian that is invariant under global $N=1$ supersymmetry by integrating this chiral superfield, or any chiral function $F(\Phi)$ of multiple chiral superfields--such as the superpotential $W$--over half of superspace. That is,
\bea
\mathcal{L} &=& \int d^2 \theta F(\Phi_{i}(x, \theta, \bar{\theta}) )+ \mathrm{h.c.}\ ,
\eea
where we have made the Lagrangian manifestly real by adding the hermitian conjugate. 
For a more general function of both chiral and anti-chiral superfields, $\mathcal{O}(\Phi, \Phi^\dagger)$, one can continue to get an $N=1$ supersymmetric invariant Lagrangian by first applying the chiral projector $-\frac{1}{4}\bar{D}^2$ to the function--hence turning it into a chiral superfield--before integrating over half of superspace,
\bea
\mathcal{L} &=&  -\frac{1}{4} \int d^2 \theta \bar{D}^2 \mathcal{O} (\Phi, \Phi^{\dagger}) + \mathrm{h.c.} \ .
\label{flatspace}
\eea

We now use similar methods to construct a Lagrangian that is invariant under local $N=1$ supersymmetry; that is, $N=1$ supergravity. To do this, one replaces the measure $d^2 \theta$ with $d^2 \Theta 2 \mathcal{E}$, where the $\Theta^\alpha$ are the covariant theta variables defined in \cite{Wess}, and $\mathcal{E}$ is a chiral density
whose lowest component is the determinant of the vierbein $e_{\mu}^{~a}$. A function of purely chiral superfields then yields the invariant Lagrangian
\bea
\mathcal{L} &=&  \int d^2 \Theta 2 \mathcal{E} F(\Phi) + \mathrm{h.c.} \ .
\eea
Once again, we can integrate over a more general function $\mathcal{O}(\Phi, \Phi^\dagger)$ by using the chiral projector. To do this, one must first replace the flat-space differential operator $\bar{D}^2$ by the covariant operator $\bar{\mathcal{D}}^2$. However, it turn out that this, by itself, is insufficient. One must also introduce a new term proportional to the chiral superfield $R$, which contains the Ricci scalar $\mathcal{R}$ in the highest component of its $\Theta$ expansion. Note that this quantity should not be confused with the $AdS_5$ radius of curvature, also denoted by $\mathcal{R}$, which will not appear explicitly in our supergravity expressions. The chiral projector in curved superspace is then given by $-\frac{1}{4}(\bar{\mathcal{D}}^2 - 8 R)$, and equation \eqref{flatspace} has as its supergravity analogue
\bea
\mathcal{L}&=&  -\frac{1}{4} \int d^2 \Theta 2 \mathcal{E} (\bar{\mathcal{D}}^2 - 8 R) \mathcal{O}(\Phi, \Phi^\dagger) + \mathrm{h.c.} \ .
\eea

\subsection{The N=1 Supergravity Galileons}
We are now ready to give the $N=1$ supergravity extension of the first three conformal Galileons. They are
\bea
\bar{\mathcal{L}}_1 &=&
\int d^2 \Theta 2\mathcal{E} 
W(\Phi)
+
\mathrm{h.c.} \ ,
\label{SUGRA-L1} 
\\
\nn \\
\bar{\mathcal{L}}_2 &=&
\M \int d^2 \Theta 2\mathcal{E} 
\bigg[
 - \frac{3}{8} ( \oD^2 - 8 R) e^{-K(\Phi, \Phi^\dagger)/3\M} 
\bigg]
+
\mathrm{h.c.} \ ,
\label{SUGRA-L2}
\\
\nn \\
\bar{\mathcal{L}}_3 &=& \frac{1}{8} \bar{\mathcal{L}}_{3, \mathrm{I}} - \frac{3}{4} \bar{\mathcal{L}}_{3, \mathrm{II}} \ ,
\eea
where
\bea
K(\Phi, \Phi^\dagger) &=&
\frac{2}{3} \frac{\mathcal{M}^4}{(\Phi + \Phi^\dagger)^2} \ ,
\\
\nn \\
\bar{\mathcal{L}}_{3, \, \mathrm{I}}   &=&
-\frac{1}{4}
\int d^2 \Theta 2\mathcal{E} ( \oD^2 - 8 R) \bigg[ (\Phi + \Phi^\dagger)^{-3} \mathcal{D} \Phi \mathcal{D} \Phi \oD^2 \Phi^\dagger \bigg] + \mathrm{h.c.}  \ ,
\label{SUGRA-L3-1}
\\
\bar{\mathcal{L}}_{3, \, \mathrm{II}} &=&
-\frac{1}{8}
\int d^2 \Theta 2\mathcal{E} ( \oD^2 - 8 R) 
\bigg[ (\Phi + \Phi^\dagger)^{-4} 
\mathcal{D} \Phi \mathcal{D} \Phi \oD \Phi^\dagger \oD \Phi^\dagger 
\bigg] 
+ \mathrm{h.c.} \ .
\label{SUGRA-L3-2}
\eea
Note that we have now restored canonical dimensions to the chiral superfields with respect to the mass scale ${\cal{M}}=1/{\cal{R}}$ of the $AdS_{5}$ bulk space. Specifically, $A$ has mass dimension 1, $F$ has dimension 2, and the superpotential $W$ has dimension 3. In addition to the $AdS_{5}$ scale ${\cal{M}}$, we have introduced the gravitational reduced Planck mass in four dimensions, denoted by $\M = 1/ (8 \pi G_N)$. All fields in the $N=1$ gravity supermultiplet, that is, the vielbein $e_{\mu}^{~a}$, the gravitino $\psi_{\mu}^{~\alpha}$ and the auxiliary fields $b_{\mu}$ and $M$ also have their mass dimensions specified with respect to ${\cal{M}}$. These are of dimension 0, 3/2, 1 and 1 respectively. At the end of this section, we will demonstrate how, for momenta much smaller than $M_{P}$, one can return to the conventions of the previous section. We also note that the expression given in \eqref{SUGRA-L3-1} and \eqref{SUGRA-L3-2} have previously been evaluated in \cite{Koehn:2013upa} - see also \cite{Ciupke:2016agp}. Finally, recall that the metric $g_{\mu\nu}=e_{\mu}^{a}e_{\nu a}$ and is dimensionless.

In component fields, we find
\bea
\frac{1}{e} \bar{\mathcal{L}}_1 &=&
- W M^* - W^* M 
+  \frac{\partial W}{\partial A} F 
+  \frac{\partial W^*}{\partial A^*} F^* \ ,
\label{A1}
\\
\frac{1}{e} \bar{\mathcal{L}}_2 &=&
\M
e^{- \tfrac{1}{3} \K} (- \tfrac{1}{2} \mathcal{R} - \tfrac{1}{3} M M^* + \tfrac{1}{3} b^\mu b_\mu) +
3 \M \frac{\partial^2 e^{-\tfrac{1}{3} \K } }{ \partial A \, \partial A^*}
\big(
\partial A \cdot \partial A 
-F F^*\big)
\label{A2}
 \\
&+&
i \M b^\mu 
\bigg(
\partial_\mu A \frac{\partial e^{-\tfrac{1}{3} \K }}{\partial A} - \partial_\mu A^* \frac{\partial e^{-\tfrac{1}{3} \K }}{\partial A^*}
\bigg)
\nn \\
&+&
\M \bigg( M F \frac{\partial e^{-\tfrac{1}{3} \K }}{\partial A}+ 
 M^* F^* \frac{\partial e^{-\tfrac{1}{3} \K }}{\partial A^*}\bigg) \ ,
\nn \\
\nn \\
\frac{1}{e}
\bar{\mathcal{L}}_{3}
&=&
\frac{1}{(A + A^*)^{3}}
\bigg[
~~(\partial A)^2 
\bigg(
2 g^{\mu \nu} \mathcal{D}_\mu \partial_\nu A^* + \tfrac{4}{3} i b^\mu \partial_\mu A^* - \tfrac{2}{3} F^* M^* 
\bigg)
\nn \\ 
&&\qquad \qquad \quad
+ (\partial A^*)^2 
\bigg(
2 g^{\mu \nu} \mathcal{D}_\mu \partial_\nu A - \tfrac{4}{3} i b^\mu \partial_\mu A - \tfrac{2}{3} F M 
\bigg)
\nn \\
&~&
\qquad \qquad \quad 
-  2 F^2 F^* M 
-  2 F^{*2} F M^*
+ 4 \, F^* ( \partial F \cdot \partial A )
+ 4 \, F ( \partial F^* \cdot \partial A^* )
\label{A3}
\\
&& \qquad \qquad \quad
+ \tfrac{1}{3} i F F^* b^\mu ( \partial_\mu A - \partial_\mu A^*)
\bigg]
\nn \\
&-& 
\frac{6}{(A + A^*)^{4}}
\bigg[
2 (\partial A )^2 (\partial A^*)^2
+ 8 \, (\partial A \cdot \partial A^* ) F F^* +  \big( (\partial A )^2 + (\partial A^* )^2  \big) F F^*
\bigg] \ ,
\nn \\
\nn 
\eea
where $e$ on the left-hand side of these expressions is the determinant of the vierbein $e_{\mu}^{~a}$--not to be confused with Euler's constant $e$ which will always appear raised to some exponent.
As in the preceding sections, we have continued to omit any terms containing the fermion $\psi$ of the chiral superfield. In addition, we also omit all interactions involving the gravitino $\psi_{\mu}^{~\alpha}$. However, we carefully analyze all terms containing the auxiliary fields; that is, $F$ which arises from the chiral superfield $\Phi$ and two new auxiliary fields. These are  $b_\mu$, a four-vector, and $M$, a complex scalar. These supergravity auxiliary fields arise in the $\Theta$ expansions of $\mathcal{E}$ and $R$. Details on how one arrives at the expressions in \eqref{A1},\eqref{A2} and  \eqref{A3} from equations \eqref{SUGRA-L1} to \eqref{SUGRA-L3-2} are given in the Appendices. Note that each of the above Lagrangians has mass dimension 4.

We can once again write out the total Lagrangian\footnote{This time limited to $\bar{\mathcal{L}}_i$, $i=1,2,3$ only.} as the sum of the individual terms given above,
\bea
\bar{\mathcal{L}} = \bar{c}_1 \bar{\mathcal{L}}_1 +  \bar{c}_2\bar{\mathcal{L}}_2 +  \bar{c}_3\bar{\mathcal{L}}_3 \ ,
\eea
where the $\bar{c}_i$'s are now dimensionless constants.
In order to restore the non-linear sigma model kinetic term in \eqref{SUGRA-L2}, we perform the following Weyl rescaling
\bea
e_\mu^{~a} \rightarrow e_\mu^{~a} ~ e^{\tfrac{1}{6} \K } \ .
\eea
This induces the following transformations:
\bea
&& \qquad e \rightarrow ~ e ~e^{\tfrac{2}{3} \K} \ , \quad g_{\mu\nu} \rightarrow g_{\mu\nu}  e^{\tfrac{1}{3} \K } \ , ~ 
\nn \
\\
&&\Gamma_{\mu \nu}^{\lambda} \rightarrow \Gamma_{\mu \nu}^{\lambda}  + \frac{1}{6M_P^2} 
\big[ \frac{\partial K}{\partial A}\big( \delta_{(\nu}^{\lambda} \partial_{\mu)} A
+ g_{\mu \nu} \partial^\lambda A 
\big)+hc. \big] \ .
\label{ss1}
\eea
For example, using \eqref{ss1}, we find that the $g^{\mu\nu} \mathcal{D}_\mu  \partial_\n A$ term in Lagrangian \eqref{A3} transforms as 
\be
g^{\mu\nu} \mathcal{D}_\mu  \partial_\n A \rightarrow 
e^{-\tfrac{1}{3} \K } g^{\mu\nu} \mathcal{D}_\mu  \partial_\nu A + 2e^{-\tfrac{1}{2} \K } \partial^\mu e^{\tfrac{1}{6} \K } \partial_\nu A \ .
\ee
We will denote the rescaled Lagrangian as $\bar{\mathcal{L'}}$, but continue to write the rescaled metric and vierbein as $g_{\mu\nu}$ and  $e_\mu^{~a}$ respectively. The Weyl rescaling restores the canonical Ricci scalar term  $- \frac{1}{2}e\M \mathcal{R}$, but also introduces a total derivative term which depends on the rescaling factor. However, since this total divergence is inside of an integral in the action, we will drop it henceforth.

We now integrate out the auxiliary fields of supergravity. We begin by first isolating the terms in the rescaled Lagrangian containing $b_\mu$. These are 
\bea
\frac{1}{e} \bar{\mathcal{L'}}_b &=&
\tfrac{1}{3} \bar{c}_2 \M b^\mu b_\mu 
+ \bar{c}_2  \M  b^\mu j_\mu 
+ \bar{c}_3 b^\mu h_\mu  \ ,
\label{eq:Lb}
\eea
where
\bea
j_\mu 
&=&
-\frac{i}{3\M} \big(\frac{\partial K}{\partial A}\partial_{\mu }A - \frac{\partial K}{\partial A^*}\partial_{\mu }A^*\big) \ ,
\nn \\
h_\mu &=& 
\frac{i}{3(A+A^*)^3} 
\big(
4  \partial_\mu A^* (\partial A)^2 + e^{\tfrac{1}{3} \K} \partial_\mu A F F^* 
- 4 \partial_\mu A (\partial A^*)^2 - e^{\tfrac{1}{3} \K} \partial_\mu A^* F F^* 
\big) \ .
\nn \\
\eea
Here, we have taken care to distinguish the ``usual" two-derivative terms from $\bar{\mathcal{L}}_2$ (collected in $j_\mu$) from the higher derivative terms (denoted by $h_\mu$) arising from $\bar{\mathcal{L}}_3$.
Solving the equation of motion for $b_\mu$ gives us
\bea
b_\mu = - \frac{3}{2} \big( j_\mu + \frac{1}{\M} \frac{\bar{c}_3}{\bar{c}_2}  h_\mu \big) \ .
\eea
Inserting this result back into Lagrangian \eqref{eq:Lb}, we find
\bea
\frac{1}{e} \bar{\mathcal{L'}}_b &=&
- \frac{1}{3} \bar{c}_2 \M b^\mu b_\mu
\nn \\
&=&
-\frac{3}{4} \bar{c}_2 \M 
\big(
j^\mu j_\mu
+ \frac{\bar{c}_3}{\bar{c}_2} \frac{2}{\M}  j^\mu h_\mu 
+ \frac{\bar{c}_3^2}{\bar{c}_2^2}\frac{1 }{M_P^4}  h^{\mu} h_\mu 
 \big)\ .
\eea
We now turn to the auxiliary field $M$, whose Lagrangian, after Weyl rescaling, is found to be
\bea
\frac{1}{e} \bar{\mathcal{L'}}_M &=&
- \bar{c}_1 e^{\tfrac{2}{3} \K}  \big( W M^* + W^* M \big)
\nn \\
&&- \tfrac{1}{3} \bar{c}_2 e^{\tfrac{1}{3} \K} M F \frac{\partial K}{\partial A}
- \tfrac{1}{3} \bar{c}_2 e^{\tfrac{1}{3} \K} M^* F^* \frac{\partial K}{\partial A^*}
- \tfrac{1}{3} \bar{c}_2 \M  e^{\tfrac{1}{3} \K} M M^* 
\nn \\
&& 
-\bar{c}_3 \frac{1}{(A + A^*)^3}
\big(
\tfrac{2}{3} e^{\tfrac{1}{3} \K}  (\partial A)^2 F^* M^* + 2 e^{\tfrac{2}{3} \K}  F^2 F^* M 
\nn \\
&& \qquad \qquad \qquad 
+\tfrac{2}{3}  e^{\tfrac{1}{3} \K}  (\partial A^*)^2 F M + 2  e^{\tfrac{2}{3} \K} (F^*)^2 F M^*  
\big) \ .
\eea
As in ordinary supergravity, we eliminate $M$ and $M^*$ by first re-writing the Lagrangian in terms of $N = M + \frac{1}{\M} \frac{\partial K}{\partial A^*} F^*$ and its complex conjugate $N^* = M^* + \frac{1}{\M} \frac{\partial K}{\partial A} F$ .
This allows one to express
\be 
\bar{\mathcal{L'}}_{M} = \bar{\mathcal{L'}}_{N }+ \bar{\mathcal{L'}}_{N_{F}}, 
\label{ss2}
\ee
where $\bar{\mathcal{L'}}_{N_{F}}$ contains terms that depend on $F, F^*$ only. 
We find that
\bea
\frac{1}{e} \bar{\mathcal{L'}}_N &=&
-\tfrac{1}{3} \bar{c}_2 \M e^{\tfrac{1}{3} \K} N N^* + N X^* + N^* X \, ,
\label{LaN1}
\\
\nn \\
\frac{1}{e} \bar{\mathcal{L'}}_{N_{F}} &=&
\bar{c}_1 e^{\tfrac{2}{3} \K} 
\big(
\frac{1}{\M}  \frac{\partial K}{\partial A} W F
+
\frac{1}{\M}  \frac{\partial K}{\partial A^*} W^* F^*
\big)
+ \bar{c}_2 \frac{1}{3\M} e^{\tfrac{1}{3} \K} \frac{\partial K}{\partial A} \frac{\partial K}{\partial A^*} F F^*
\nn \\
&+&
\bar{c}_3\frac{1}{(A + A^*)^3} \frac{1}{\M} 
\big[
\tfrac{2}{3}  e^{\tfrac{1}{3} \K} \frac{\partial K}{\partial A} (\partial A)^2 F F^*
+
2 e^{\tfrac{2}{3} \K} \frac{\partial K}{\partial A^*} (F F^*)^2
\nn \\
&& \qquad \qquad \qquad \quad  
+
\tfrac{2}{3}  e^{\tfrac{1}{3} \K} \frac{\partial K}{\partial A^*} (\partial A^*)^2 F F^*
+
2  e^{\tfrac{2}{3} \K} \frac{\partial K}{\partial A} (F F^*)^2
\big] \ ,
\nn \\
\label{La-extra}
\eea
where
\bea
X &=&
-\bar{c}_1 e^{\tfrac{2}{3}\K} W 
- 
\bar{c}_3 \frac{1}{(A+A^*)^3}
\big[
\tfrac{2}{3}e^{\tfrac{1}{3}\K} (\partial A)^2 F^* 
+ 2 e^{\tfrac{2}{3} \K} (F^*)^2 F 
\big] \ .
\eea
The equation of motion for $N$ is straightforward to solve. The solution is
\bea
N &=& \frac{1}{\bar{c}_2} \frac{3}{ \M }  e^{-\tfrac{1}{3} \K} X \ .
\eea
Substituting this result into \eqref{LaN1}, we find
\bea
\frac{1}{e} \bar{\mathcal{L'}}_N &=& \tfrac{1}{3} \bar{c}_2 \M e^{-\tfrac{1}{3} \K} N N^*
\nn \\
&=&
 \frac{3}{\M} 
\bigg[
\frac{\bar{c}_1^{\, 2}}{\bar{c}_2}e^{ \K} W W^*
\nn \\
&+& 
\frac{\bar{c}_1 \bar{c}_3}{\bar{c}_2} \frac{1}{(A +A^*)^3}  
\bigg(
\tfrac{2}{3} e^{\tfrac{2}{3} \K}\big[ (\partial A^*)^2 FW + (\partial A)^2 F^*W^*\big] 
+ 2 e^{\K} \big[F^2 F^*W  + (F^*)^2 FW^* \big]
\bigg)
\nn \\
&+&  
\frac{\bar{c}_3^{\, 2}}{\bar{c}_2}  \frac{1}{(A +A^*)^6} 
\bigg(
 \tfrac{4}{9}  e^{\tfrac{1}{3} \K} (\partial A)^2 (\partial A^*)^2 F^* F
+ \tfrac{4}{3}e^{\tfrac{2}{3}\K} \big[(\partial A)^2 + (\partial A^*)^2 \big](F^*F)^2 
+4  e^{\K} (F^* F)^3
\bigg)
\bigg] 
\nn \\
\eea
Combining $\bar{\mathcal{L'}}_b$ , $\bar{\mathcal{L'}}_N$ and $\bar{\mathcal{L'}}_{N_{F}}$ with the remaining terms in $\bar{\mathcal{L'}}$, we arrive at the complete Lagrangian. It is given by
\bea
\dfrac{1}{e}\bar{\mathcal{L'}} &=&\bar{c}_1 e^{\tfrac{2}{3} \K}
\big(
\frac{\partial W }{\partial A } F + \frac{\partial W^* }{\partial A^* } F^* 
+  
\frac{1}{\M} \frac{\partial K}{\partial A} W F + \frac{1}{\M} \frac{\partial K}{\partial A^*} W^* F^*
\big)
+ \frac{\bar{c}_1^{\, 2}}{\bar{c}_2}\frac{3}{\M}  e^{\K} W W^*
\nn \\
\nn \\
&-&\bar{c}_2\frac{\M}{2}\mathcal{R} 
- \bar{c}_2\frac{\partial^2 K}{\partial A \partial A^*} \partial A \cdot \partial A^*
+ \bar{c}_2 e^{\tfrac{1}{3} \K } \frac{\partial^2 K}{\partial A \partial A^*} F F^*
\nn \\
&-& \frac{3}{4} \M e^{\tfrac{2}{3} \K} 
\big(
\frac{\bar{c}_3}{\bar{c}_2}\frac{2 }{\M}  j^\mu h_\mu + \frac{\bar{c}_3^{\,2}}{\bar{c}_2^{\, 2}} \frac{1 }{M_P^4}   h^{\mu} h_\mu
\big)
\nn \\
\nn \\
&+&
\bar{c}_3
\frac{1}{(A+ A^*)^3}
\big(
~~(\partial A)^2 
\big(
2 g^{\mu \nu} \mathcal{D}_\mu \partial_\nu A^* 
+\frac{1}{3 \M}
\big(
\frac{\partial K}{\partial A} \partial A \cdot \partial A^*
+ 
\frac{\partial K}{\partial A} (\partial A^*)^2
\big)
\big)
\nn \\
&& \qquad \qquad \qquad
+ (\partial A^*)^2 
\big(
2 g^{\mu \nu} \mathcal{D}_\mu \partial_\nu A  
+\frac{1}{3 \M}
\big(
\frac{\partial K}{\partial A} (\partial A)^2
+ 
\frac{\partial K}{\partial A} \partial A^* \cdot \partial A
\big)
\big)
\nn \\
&~&
\qquad \qquad \qquad 
+ 4  e^{\tfrac{1}{3} \K}F^* ( \partial F \cdot \partial A )
+ 4  e^{\tfrac{1}{3} \K}F ( \partial F^* \cdot \partial A^* )
\big)
\nn \\
&-& 
\bar{c}_3\frac{6}{(A + A^*)^{4}}
\big(
2 (\partial A )^2 (\partial A^*)^2
+ 8 e^{\tfrac{1}{3} \K} (\partial A \cdot \partial A^* ) F F^* +  e^{\tfrac{1}{3} \K} \big( (\partial A )^2 + (\partial A^* )^2  \big) F F^*
\big)
\nn\\
\nn \\
&+&  
\frac{\bar{c}_1 \bar{c}_3}{\bar{c}_2} \frac{1}{(A +A^*)^3} \frac{3}{\M}  
\big(
\tfrac{2}{3} e^{\tfrac{2}{3} \K}\big[ (\partial A^*)^2 FW + (\partial A)^2 F^*W^*\big] 
+ 2 e^{\K} \big[F^2 F^*W  + (F^*)^2 FW^* \big]
\big)
\nn \\
&+&  
\frac{\bar{c}_3^{\, 2}}{\bar{c}_2}  \frac{1}{(A +A^*)^6} \frac{3}{\M} 
\big(
 \tfrac{4}{9}  e^{\tfrac{1}{3} \K} (\partial A)^2 (\partial A^*)^2 F^* F
+ \tfrac{4}{3}e^{\tfrac{2}{3}\K} \big[(\partial A)^2 + (\partial A^*)^2 \big](F^*F)^2 
+4  e^{\K} (F^* F)^3
\big)
\nn \\
\nn \\
&+&
\bar{c}_3\frac{1}{(A + A^*)^3} \frac{1}{\M}
\big(
\tfrac{2}{3}  e^{\tfrac{1}{3} \K} \big( \frac{\partial K}{\partial A} (\partial A)^2  + \frac{\partial K}{\partial A^*} (\partial A^*)^2 \big)F^* F
+
2 e^{\tfrac{2}{3} \K} \big( \frac{\partial K}{\partial A} + \frac{\partial K}{\partial A^*} \big) (F^* F)^2
\big) \ ,
\nn \\
\label{eq:L_wF}
\eea
where
\bea
 j^\mu h_\mu &=&
\frac{1}{(A + A^*)^3} \frac{1}{9 \M}
\bigg[
4 \frac{\partial K}{\partial A} (\partial A \cdot \partial A^*) (\partial A)^2 
- 
4 \frac{\partial K}{\partial A}  (\partial A)^2 (\partial A^*)^2 
\nn \\
&& \qquad \qquad \quad \quad~
- 
4\frac{\partial K}{\partial A^*} ( \partial A^*)^2 (\partial A)^2 
+
4\frac{\partial K}{\partial A^*}  (\partial A \cdot \partial A^*) (\partial A^*)^2 
\nn \\
&&\qquad \qquad \quad \quad~
+ 
e^{\tfrac{1}{3} \K}\frac{\partial K}{\partial A}(\partial A)^2 F F^* 
-
e^{\tfrac{1}{3} \K}\frac{\partial K}{\partial A} (\partial A \cdot \partial A^*) F F^* 
\nn \\
&& \qquad \qquad \quad \quad~
-
e^{\tfrac{1}{3} \K}\frac{\partial K}{\partial A^*}(\partial A^* \cdot \partial A)  F F^*
+
e^{\tfrac{1}{3} \K} \frac{\partial K}{\partial A^*} ( \partial A^*)^2 F F^* 
\bigg] ,
\nn \\
\eea
\bea
h^\mu h_\mu 
=
&\dfrac{-1}{9 (A + A^*)^6}&
\bigg[
16 (\partial A^*)^2 (\partial A )^4
-
32  \partial A \cdot \partial A^* (\partial A )^2(\partial A^*)^2
+
16  (\partial A)^2 (\partial A^*)^4
\nn \\
&\quad\quad\quad&
+ 
16 e^{\tfrac{1}{3} \K } (\partial A^* \cdot \partial A )(\partial A)^2 F F^*
-
16 e^{\tfrac{1}{3} \K } (\partial A^*)^2 (\partial A)^2 F F^*
\nn \\
&\quad \quad \quad&
+
e^{\tfrac{2}{3} \K} (\partial A)^2 (F F^*)^2
-
 2e^{\tfrac{2}{3} \K}\partial A \cdot \partial A^* (F F^*)^2
+
e^{\tfrac{2}{3} \K }(\partial A^*)^2 (FF^*)^2
\bigg] ,
\nn \\
\eea
and 
\bea
K = K(A, A^*) = \frac{2}{3} \frac{\mathcal{M}^4}{(A +A^*)^2} \ .
\eea
We want to emphasize that the final result \eqref{eq:L_wF} is exact, and has not used the $\partial^{2},F \ll 1$ limit employed in the previous section.

As an important check on our supergravitational expression for $\mathcal{\bar{L}'}$ in \eqref{eq:L_wF}, let us take the flat superspace limit. To do this, we let $\M \rightarrow \infty$ and $g_{\mu \nu} \rightarrow \eta_{\mu \nu}$, $e \rightarrow 1$. We find that
\bea
\mathcal{\bar{L}'} &=&
\bar{c}_1 
\big(
\frac{\partial W }{\partial A } F + \frac{\partial W^* }{\partial A^* } F^* 
\big)
- \bar{c}_2\frac{\partial^2 K}{\partial A \partial A^*} \partial A \cdot \partial A^*
+ \bar{c}_2  \frac{\partial^2 K}{\partial A \partial A^*} F F^*
\nn \\
&+&
\bar{c}_3
\frac{1}{(A+ A^*)^3}
\big[
2 (\partial A)^2 \square A^* + 2 (\partial A^*)^2 \square A
+ 4 F^* (\partial F \cdot \partial A) + 4 F (\partial F^* \cdot \partial A^*)
\big] 
\nn \\
&-&
\bar{c}_3 \frac{6}{(A+ A^*)^4}
\big[
2(\partial A)^2 (\partial A^*)^2 
+  8(\partial A \cdot \partial A^*) F F^*
+ \big( (\partial A)^2 + (\partial A^*)^2 \big) F F^* 
\big] \ .
\nn \\
\label{SUGRA-flat-limit}
\eea
After the following integration by parts,
\bea
 \frac{2}{(A + A^*)^3} \big( F^* \partial F \cdot \partial A + F \partial F^* \cdot \partial A^* \big)
&=& 
-\frac{2}{(A + A^*)^3} \big( F^* F \square (A + A^*) +  F \partial F^* \partial A + F^* \partial F \partial A^*  \big)
\nn \\
&&
+\frac{6}{(A + A^*)^4} \big( (\partial A)^2 + (\partial A^*)^2 + \partial A \cdot \partial A^* \big)F^* F \ ,
\nn \\
\eea
equation \eqref{SUGRA-flat-limit} is the sum of the flat superspace conformal Galileons of the previous section--now, however, written in terms of $A=\frac{1}{\sqrt{2}}(\phi+i\chi)$ and expressed in the canonical mass scale conventions defined near the beginning of this section.
To return to the field normalization conventions used in the previous section of the paper, we implement the following procedure.
We use the mass scale $\mathcal{M}$ to set $A \rightarrow \mathcal{M} A$, $F \rightarrow \mathcal{M} F$, $W \rightarrow \mathcal{M}^3 W$ and restore the dimensions in the constants by letting $\bar{c}_i \rightarrow \bar{c}_i / \mathcal{M}^4$. 
For example, 
\bea
- \bar{c}_2\frac{ 4 \mathcal{M}^4 }{(A + A^*)^4} \partial A \cdot \partial A^*
\rightarrow
- \frac{\bar{c}_2}{\mathcal{M}^4 }\frac{4 \mathcal{M}^4 }{\mathcal{M}^4 (A + A^*)^4 }  \mathcal{M}^2 \partial A \cdot \partial A^*
= 
- \bar{c}_2\frac{4 }{(A + A^*)^4} \bigg( \frac{\partial A}{\mathcal{M}} \bigg) \cdot \bigg( \frac{\partial A^*}{\mathcal{M}} \bigg) \ ,
\eea
where $A$ is now dimensionless while $\bar{c}_2$ has dimension 4. Similarly, one finds
\bea
&&
\bar{c}_1 \frac{\partial W}{\partial A} F
\rightarrow
\frac{\bar{c}_1}{\mathcal{M}^4} \frac{\mathcal{M}^3 \partial W}{\mathcal{M} \partial A} \mathcal{M}F
=
\bar{c}_1 \frac{\partial W}{\partial A}  \frac{F}{{\cal{M}}}
\nn \\
&&
\bar{c}_3
\frac{1}{(A+ A^*)^3} 2 (\partial A)^2 \square A^* 
\rightarrow
\frac{\bar{c}_3}{\mathcal{M}^4}
\frac{1}{\mathcal{M}^3 (A+ A^*)^3} 2 \mathcal{M}^3(\partial A)^2 \square A^* 
=
\bar{c}_3
\frac{1}{ (A+ A^*)^3} 2 \bigg(\frac{\partial A}{\mathcal{M}} \bigg)^2 \frac{\square A^*}{\mathcal{M}^2}, 
\nn \\
\eea
where $W$ has mass dimension 0, $F$ has dimension 1, and $\bar{c}_1, \bar{c}_3$ have dimension 4. Continuing this procedure and again setting ${\cal{M}}=1$, the flat superspace limit of \eqref{eq:L_wF} yields, as expected,
\bea
\mathcal{\bar{L}'} &=&
\bar{c}_1 
\big(
\frac{\partial W }{\partial A } F + \frac{\partial W^* }{\partial A^* } F^*  
\big)
+\bar{c}_2\big(-\frac{\partial^2 K}{\partial A \partial A^*}  \partial A \cdot \partial A^*
+  \frac{\partial^2 K}{\partial A \partial A^*} F F^*)
\nn \\
&+&
\bar{c}_3\bigg[
\frac{1}{(A+ A^*)^3}
\big[
2 \big(\partial A \big)^2 \square A^*
+ 2 \big(\partial A^* \big)^2 \square A
- 2 F^* F \square ( A + A^*)
\nn \\
&&\qquad \qquad \qquad
+ 2 F^* \partial F \cdot \partial (A - A^*)
- 2 F \partial F^* \cdot \partial (A - A^*)
\big] 
\nn \\
&-&
\frac{6}{(A+ A^*)^4}
\big[
2(\partial A)^2 (\partial A^*)^2
+  6 (\partial A \cdot \partial A^*) F F^*
\big]\bigg] \ .
\nn \\
\label{SUGRA-flat-limit-2}
\eea
This is precisely the sum of the dimensionless flat superspace conformal Galileons defined in equations (\ref{33}), (\ref{bw1}), and (\ref{bw2}) written in terms of $A=\frac{1}{\sqrt{2}}(\phi+i\chi)$.

\subsection{The Low Momentum, Curved Spacetime Limit}

As mentioned above, expression \eqref{eq:L_wF} is exact and is valid for any momentum or magnitude of $F$. However, one important application of \eqref{eq:L_wF} is in a cosmological context where 1) ${\cal{M}} \ll M_{P}$, 2) although spacetime can be curved and dynamical, its curvature $\mathcal{R} \ll M_{P}^{2}$ and 3) the momentum and auxiliary field $F$ of chiral matter also satisfy $\partial^{2}, F \ll M_{P}^{2}$. Be this as it may, 4) it is not necessary for $\partial^{2}$ or $F$ to be smaller than ${\cal{M}}^{2}$. In this ``cosmological'' limit, the expression \eqref{eq:L_wF} greatly simplifies. Neglecting all terms--with the notable exception of $\frac{\M}{2}\mathcal{R}$--that depend explicitly on $M_{P}$, the action associated with \eqref{eq:L_wF} becomes
\be
S=\int{\sqrt{-det g}~ \mathcal{\bar{L}'}_{\rm cosmo}}
\label{bw4} 
\ee
where 
\bea
\mathcal{\bar{L}'}_{\rm cosmo} &=& 
-{\bar{c}}_{2}\frac{\M}{2}\mathcal{R} +\bar{c}_1 
\big(
\frac{\partial W }{\partial A } F + \frac{\partial W^* }{\partial A^* } F^*  
\big)
+\bar{c}_2\big(-\frac{\partial^2 K}{\partial A \partial A^*}  \nabla A \cdot \nabla A^*
+  \frac{\partial^2 K}{\partial A \partial A^*} F F^*)
\nn \\
&+&
\bar{c}_3\bigg[
\frac{1}{(A+ A^*)^3}
\big[
2 \big(\nabla A \big)^2 \nabla^{2} A^*
+ 2 \big(\nabla A^* \big)^2 \nabla^{2} A
- 2 F^* F \nabla^{2} ( A + A^*)
\nn \\
&&\qquad \qquad \qquad
+ 2 F^* \nabla F \cdot \nabla (A - A^*)
- 2 F \nabla F^* \cdot \nabla (A - A^*)
\big] 
\nn \\
&-&
\frac{6}{(A+ A^*)^4}
\big[
2(\nabla A)^2 (\nabla A^*)^2
+  6 (\nabla A \cdot \nabla A^*) F F^*
\big]\bigg] \ .
\label{bw5}
\eea
All indices in this expression are contracted with respect to a curved spacetime metric $g_{\mu\nu}$ and $\nabla$ is the associated covariant derivative.

\subsection{Low Momentum, Curved Spacetime Limit--Including ${\bar{\cal{L}}}^{SUSY}_{4}$}

Similarly to the above discussion of ${\bar{\cal{L}}}_{3}^{SUGRA}$, we have also carried out a complete calculation of the supergravity Lagrangian ${\bar{\cal{L}}}_{4}^{SUGRA}$. This is a complicated and detailed analysis and will be presented elsewhere. However, the results of that calculation that concern the low-energy ``cosmological'' limit defined above are relatively straightforward. We find that, as for ${\bar{\cal{L}}}_{3}^{SUGRA}$, in this limit all terms in the expression \eqref{bw7} contribute to the low-energy curved superspce Lagrangian --however, with the modification that one replaces the flat metric $\eta_{\mu\nu}$ with the curved spacetime metric $g_{\mu\nu}$ and each derivative with the covariant derivative with respect to $g_{\mu\nu}$. It follows that to \eqref{bw5} one must add the term
\bea
&\bar{c}_{4}& \bigg[ 
\frac{1}{(A+A^{*})^{2}}\bigg[2\nabla_{\mu}(\nabla A)^{2} \nabla^{\mu}(\nabla{A}^{*})^{2} 
-4|\nabla A |^{2} |\nabla F|^{2}  +12|F|^{2}|\nabla F|^{2}
\nn \\
&+&
\bigg( \big(
2 \nabla^{\mu}F {\nabla}^{2} A  + \nabla^{\mu} (\nabla A \cdot \nabla F) 
\big) F^* +\big(2 \nabla^{\mu}F^* {\nabla}^{2} A^*  + \nabla^{\mu} (\nabla A^* \cdot \nabla F^*) 
\big) F \bigg)\nabla_{\mu}(A + A^*)
\nn \\
\nn \\
&+&\big(-6 \nabla_{\mu} \nabla_{\nu}A\nabla^{\mu}\nabla^{\nu}A^{*} 
+  
\nabla^{\mu}\nabla^{\nu}A \nabla_{\mu}\nabla_{\nu} A 
+\nabla^{\mu}\nabla^{\nu}A^{*} \nabla_{\mu}\nabla_{\nu} A^{*} 
\big) |F|^{2} 
\nonumber \\
\nn \\ 
&-&  
 \nabla^{\mu }((\nabla A)^2 ) \nabla_{\mu}F F^* 
-  \nabla^{\mu }((\nabla A^*)^2 ) \nabla_{\mu}F^* F 
\nn \\
\nn \\
&-&
3\big( \nabla_{\nu}A\nabla^{\nu}\nabla^{\mu}A^{*}(\nabla_{\mu}F)F^{*} 
+\nabla_{\nu}A^{*}\nabla^{\nu}\nabla^{\mu}A F\nabla_{\mu}F^{*}\big)
\nonumber \\
\nn \\
&-&
\nabla_{\mu}\nabla_{\nu}A \nabla^{\nu}A \nabla^{\mu}FF^* 
-\nabla_{\mu} \nabla_{\nu}A^{*} \nabla^{\nu}A^{*} \nabla^{\mu}F^*F 
\label{ABC}
\\
&+& \bigg(\big(
\nabla_{\nu}F^* \nabla^ {\nu}\nabla^{\mu}A F + \nabla_{\nu}F^* \nabla^{\nu}A \nabla^{\mu}F
\big) 
+ \big(
\nabla_{\nu}F \nabla^ {\nu}\nabla^{\mu}A^{*} F^{*} + \nabla_{\nu}F \nabla^{\nu}A^{*} \nabla^{\mu}F^{*}
\big) 
\nn \\
&-&  
\nabla^{\mu}(\nabla A)^2 {\nabla}^{2} A^*
- \nabla^{\mu}(\nabla A^*)^2 {\nabla}^{2} A \Bigg)\nabla_{\mu}( A + A^*)
\bigg]
\nn \\
\nonumber
&+&
\frac{2}{(A + A^*)^{3}} \bigg[
\bigg( \big(
 \nabla^\mu (\nabla A)^2 +  \nabla^\mu(\nabla A^*)^2 
- 2 \nabla^{\mu}F F^* - 2 \nabla^{\mu}F^{*} F\big)|F|^{2}
 \\
&-&
( \nabla^{\nu} \nabla^{ \mu}A F + \nabla^{\nu}A \nabla^{\mu} F) F^* 
+( \nabla^{\nu} \nabla^{ \mu}A^* F^* + \nabla^{\nu}A^* \nabla^{\mu} F^*) F 
\bigg) \nabla_{\nu}(A - A^*)\nabla_{\mu}(A + A^*)
\nn \\
\nn \\
&+&\bigg((\nabla A)^{2}(\nabla A^{*})^{2}-2(\nabla A \cdot \nabla A^{*})|F|^{2}+ |F|^{4}\bigg){\nabla}^{2} (A+A^{*}) \bigg]
 \bigg]
\nn \\
\nonumber
\eea

However, unlike the case of ${\bar{\cal{L}}}_{3}^{SUGRA}$, there are several other terms involving the curvature tensor that also enter the ``cosmological'' limit of  ${\bar{\cal{L}}}_{4}^{SUGRA}$. The origin of these terms is less straightforward and, in this paper, we will simply state the results. We find that in, addition to \eqref{ABC}, one needs to add the terms\\
\bea
{\bar{c}}_{4}\bigg[ \frac{1 }{ 128 (A + A^*)^2 }
\bigg(
34 \mathcal{R}  \nabla_\mu (A + A^*) \nabla^{\mu} ( A  + A^*)
-9 \mathcal{R}_{\mu \nu} \nabla^\mu (A + A^*) \nabla^{\nu} ( A  + A^*)
\bigg)|F|^{2} \bigg]
\label{bw102}
\eea
to ${\bar{{\cal{L}}}}^{\prime}_{\rm cosmo}$ in \eqref{bw5}. $\mathcal{R}$ and $\mathcal{R}_{\mu\nu}$ are the spacetime curvature scalar and Ricci tensor respectively. Note that all contractions in \eqref{ABC} and \eqref{bw102} are with respect to the curved metric $g_{\mu\nu}$. Here, we will simply point out that these new terms arise from commuting certain derivatives in the supergravity extension of ${\bar{\cal{L}}}^{SUSY}_{4, \, \rm{2nd \, term}}$ in \eqref{30}. 

Finally, we note that various powers of $F$ and $\partial F$ occur in $\mathcal{\bar{L}'}_{\rm cosmo}$  given as the sum of \eqref{bw5}, \eqref{ABC} and \eqref{bw102}. The field $F$ can be replaced by $F=F^{(0)}+F^{(1)}+F^{(2)}$ in the terms proportional to ${\bar{c}}_{1}$, ${\bar{c}}_{2}$, by $F=F^{(0)}+F^{(1)}$ in the ${\bar{c}}_{3}$ terms and by $F^{(0)}$ in the terms proportional to ${\bar{c}}_{4}$. The expressions for $F^{(0)}$, $F^{(1)}$ and $F^{(2)}$ are given in \eqref{46}, \eqref{r8} and \eqref{ho2} respectively.

%%%%%%%%%%%%%%
\section*{Acknowledgments}
R. Deen and B.A. Ovrut are supported in part by DOE contract No. $\mathrm{DE}$-$\rm{SC}0007901$. B. Ovrut would like to thank Anna Ijjas , Paul Steinhardt and other members of the PCTS working group ``Rethinking Cosmology'' for many helpful conversations. Ovrut would also like to thank his long term collaborator Jean-Luc Lehners for his joint work on higher-derivative supersymmetry and supergravitation. Finally, R. Deen is grateful to Anna Ijjas and Paul Steinhardt for many discussions and the Center for Particle Cosmology at the University of Pennsylvania for their support.
%%%%%%%%%%%%%%%%%%%%%%%%%%%%%%%%%%%%%%%%%%%%%%%%

\section*{Appendix A: Constructing Higher-Derivative SUGRA Lagrangians}\label{app-A}
We give a brief explanation of how the supergravity Lagrangians in (\ref{SUGRA-L1})-(\ref{SUGRA-L3-2}), which are written in terms of superfields, can be expressed in component fields. The formalism used here is based on work presented in  \cite{Koehn:2012ar} ,  \cite{Baumann:2011nm, Farakos:2012je, Farakos:2012qu} and \cite{Wess}.
Recall that a chiral superfield $\Phi$ has the following $\Theta$ expansion
\bea
\Phi = A + \sqrt{2} \Theta \psi + \Theta \Theta F \ .
\eea
The components of $\Phi$ can be obtained by acting with $\D$ and then taking the lowest component, which we denote by ``\big |''.
For example,
\bea
F = - \tfrac{1}{4} \D^2 \Phi \big|,
\label{Theta2}
\eea
is the $\Theta^2$ component of $\Phi$. 

Within the context of $N=1$ supergravity, we are interested in constructing invariant superfield Lagrangians. This can be accomplished as follows. An integral over chiral superspace, $\int d^2  \Theta \, \mathcal{E} X$, requires the integrand $X$ to be a chiral superfield. Multiplication by the chiral density $\mathcal{E}$ means that under local supersymmetry, the entire integral transforms into a total space derivative. 
The product $\mathcal{E} X$ continues to be chiral and has an exact expansion in the local superspace coordinate $\Theta^\alpha$. 
As explained above, we can construct a chiral superfield $X$ out of any Lorentz scalar $\mathcal{O}$ by acting on it with the chiral projector $\oD^2 - 8R$.
The integral $\int d^2 \Theta \mathcal{E} X$ then projects out the $\Theta^2$ component of $\mathcal{E} X$. However, we have seen in \eqref{Theta2} that the $\Theta^2$ component of a chiral superfield can be obtained by first acting with $-\frac{1}{4}\D^2$ and then taking the lowest component.
Choosing $X = (\oD^2 - 8R) \mathcal{O}$, it follows that
\bea
\int d^2 \Theta \, 
\mathcal{E}
(\oD^2 - 8 R) \mathcal{O}
&=&
- \frac{1}{4} \D^2 \big( \mathcal{E} (\oD^2 - 8R) \mathcal{O} \big) \big| \ .
\nn \\
\eea
Under the assumption that we ignore all fermions, including the gravitino, this can be written as
\bea
\int d^2 \Theta \, 
\mathcal{E}
(\oD^2 - 8 R) \mathcal{O}
&=&
- \frac{1}{4} \mathcal{E} \big|~ \D^2 \big( (\oD^2 - 8R) \mathcal{O} \big) \big| 
- \frac{1}{4} \D^2  \mathcal{E} \big|~ \big((\oD^2 - 8R) \mathcal{O} \big) \big| 
\nn \\
&=&
- \frac{1}{4} \mathcal{E} \big|~ \D^2 \big( (\oD^2 - 8R) \mathcal{O} \big) \big| 
+ \mathcal{E} \big|_{\Theta^2}~ (\oD^2 - 8R) \mathcal{O} \big) \big| \ ,
\eea
where
\bea
R &=&
-\tfrac{1}{6}M 
+\Theta^2 
(\tfrac{1}{12}\mathcal{R} - \tfrac{1}{9}M M^* - \tfrac{1}{18}b^\mu b_\mu + \tfrac{1}{6}i e_a^{~\mu} \D_\mu b^a)
\nn \\
\mathcal{E} &=& \tfrac{1}{2}e    -\tfrac{1}{2} \Theta^2e M^*
\label{bw3}
\eea
It follows that one can compute the component field  expansion of a supergravity Lagrangian by evaluating the following terms,
\bea
\D^2 \oD^2 \mathcal{O}\big| \ , 
- 8 \D^2 (R\mathcal{O} )  \big| \ ,
\oD^2 \mathcal{O} \big| \ ,
- 8R\mathcal{O} \big| \ .
\label{m1}
\eea

As an example, consider the first term \eqref{SUGRA-L3-1} in the curved superspace $\mathcal{\bar{L}}_3$ conformal Galileon. It was constructed using the formalism just described. That is, one begins with the higher-derivative superfield expression
\be
\mathcal{O}_{\mathrm{I}} 
=(\Phi +  \Phi^\dagger)^{-3}
\D \Phi  \D \Phi  \oD^2 \Phi^\dagger \ .
\label{e1}
\ee
Then, the associated Lagrangian is obtained by the appropriate chiral projection and superspace integration. 
For $\mathcal{O}_{\mathrm{I}}$, which is not hermitian, one writes
\bea
\mathcal{\bar{L}}_{3, \mathrm{I}} 
=
- \frac{1}{4}
\int d^2 \Theta \, 
2 \mathcal{E}
(\oD^2 - 8 R) \mathcal{O}_{\mathrm{I}} 
+
\mathrm{h.c.} 
\eea
Having obtained the superfield expression for the $\mathcal{\bar{L}}_{3, \mathrm{I}}$ Lagrangian, we now apply the preceding formalism to express it in terms of component fields.
It follows from the above that one must evaluate the four lowest component terms in \eqref{m1}. To exhibit our methods, let us compute $\oD^2 \mathcal{O} \big|$.
\bea
\oD^2 \big( (\Phi +  \Phi^\dagger)^{-3}
\D \Phi  \D \Phi  \oD^2 \Phi^\dagger \big)  \big| 
&=&
\oD^2 \big( (\Phi +  \Phi^\dagger)^{-3} \big) \big|~
\D \Phi  \D \Phi  \oD^2 \Phi^\dagger  \big| 
+
2\D^{\alpha}\big( (\Phi +  \Phi^\dagger)^{-3} \big) \big|~  \D_{\alpha} \big( \D \Phi  \D \Phi  \oD^2 \Phi^\dagger \big)  \big|~ 
\nn \\
&+&
 (\Phi +  \Phi^\dagger)^{-3} \big| ~
\oD^2 \big( \D \Phi  \D \Phi  \oD^2 \Phi^\dagger \big)   \big| \ .
\label{f1}
\eea
We calculate these terms by distributing the $\D_{\alpha}$ and $\oD_{\dot{\alpha}}$ operators appropriately, and commuting them until we are able to apply the defining expressions for chiral and anti-chiral fields
\bea
\oD_{\dot{\alpha}} \Phi = 0 \ ,
\quad
\D_{\alpha} \Phi^\dagger = 0 \ . 
\eea
Many terms that arise in the intermediate stages of the calculation involve fermions. For example, expressions which contain $\D_{\alpha} \Phi \big| = \sqrt{2} \psi_\alpha$ are fermionic. In keeping with the thesis of this paper, all such terms will be dropped. 
However, the essential difficulty involved in the computation is the presence of curvature and torsion in supergravity. Hence, anti-commutators of the $\D, \oD$ operators now give rise to terms which would not have been present in the global supersymmetric case.
Explicitly, we have
\bea
\left(
\D_C \D_B - (-)^{bc} \D_B \D_C
\right) V^A 
&=&
(-)^{d(c+b)} V^D R_{CBD}^{~~~~~A} - T_{CB}^{~~D} \D_D V^A \nn \\ 
&=&
\left(
V^e R_{CBe}^{~~~~~A} + (-)^{(c+b)} V^\delta R_{CB\delta}^{~~~~~A} + (-)^{(c+b)} V_{\dot{\delta}}  R_{CB}^{~~~~\dot{\delta}A}
\right) \nn \\
&&
- \left(
T_{CB}^{~~e} \D_e V^A + T_{CB}^{~~\delta} \D_\delta V^A +  T_{CB \dot{\delta}} \oD^{\dot{\delta}} V^A
\right) \ ,
\label{commutator}
\nn \\
\eea
where the $A, B, C, D$ indices can be $a, \alpha, \dot{\alpha}$, and the exponents $b, c, d$ take the values $0$ or  $1$ when the indices $B, C, D$ are bosonic or fermionic  respectively. $R_{CBD}^{~~~~~A}$ and $T_{CB}^{~~D}$ are superfields which respectively contain components of the curvature and torsion. For $N=1$ supergravity, these superfields and their component expansions are given, for example, in \cite{Wess}, Chapter 15. 

Using these results, we determine that the first two terms in \eqref{f1} are fermionic and, hence, are taken to vanish. The third term is given by
\be
 (\Phi +  \Phi^\dagger)^{-3} \big| ~
\oD^2 \big( \D \Phi  \D \Phi  \oD^2 \Phi^\dagger \big)   \big| 
=
 (A +  A^*)^{-3} ~
\oD^2 \big( \D \Phi  \D \Phi  \oD^2 \Phi^\dagger \big)   \big| \ .
 \label{f2}
\ee
We compute the lowest component term on the right-hand-side as follows.
\bea
\noindent \oD^2 \big( \D \Phi  \D \Phi  \oD^2 \Phi^\dagger \big)   \big| &=&
\epsilon^{\dot{\beta}\dot{\alpha}} 
\oD_{\dot{\beta}} \oD_{\dot{\alpha}}(\D \Phi \D \Phi \overbar{\D}^2 \Phi^\dagger )\big| ~
\nn \\
&=&
\left(
\overbar{\D}_{\dot{\alpha}} \D \Phi \big|~ \overbar{\D}_{\dot{\beta}}\D \Phi \big|~ \overbar{\D}^2 \Phi^\dagger \big|~ 
-
\overbar{\D}_{\dot{\beta}}  \D \Phi \big|~ \overbar{\D}_{\dot{\alpha}}  \D \Phi \big|~ \overbar{\D}^2 \Phi^\dagger \big|~
\right) \nn \\
&=&
\epsilon^{\dot{\beta}\dot{\alpha}} \epsilon^{\beta \alpha}
\left(
\overbar{\D}_{\dot{\alpha}} \D_\alpha \Phi \big|~ \overbar{\D}_{\dot{\beta}}\D_\beta \Phi \big|~ 
-
\overbar{\D}_{\dot{\beta}}  \D_\alpha \Phi \big|~ \overbar{\D}_{\dot{\alpha}}  \D_\beta \Phi \big|~ 
\right) \overbar{\D}^2 \Phi^\dagger \big|~ \nn \\
&=&
\epsilon^{\dot{\beta}\dot{\alpha}} \epsilon^{\beta \alpha}
\bigg(
(-2i \sigma_{\alpha \dot{\alpha}}^{a} e_a^{~\mu} \partial_\mu A ) (-2i \sigma_{\beta \dot{\beta}}^{b} e_b^{~\nu} \partial_\n A)
-
(-2i \sigma_{\alpha \dot{\beta}}^{a} e_a^{~\mu} \partial_\mu A)(-2i \sigma_{\beta \dot{\alpha}}^{b} e_b^{~\nu} \partial_\nu A)
\bigg) (-4 F^*) 
\nn \\
&=&
16
\epsilon^{\dot{\beta}\dot{\alpha}} \epsilon^{\beta \alpha}
\left(
 \sigma_{\alpha \dot{\alpha}}^{a} \sigma_{\beta \dot{\beta}}^{b}
-
 \sigma_{\alpha \dot{\beta}}^{a} \sigma_{\beta \dot{\alpha}}^{b}
\right)  e_a^{~\mu} e_b^{~\nu}  \partial_\mu A \,  \partial_\nu A \, F^*
\nn \\
&=&
16
\left(
 \overbar{\sigma}^{a \dot{\beta}\beta} \sigma_{\beta \dot{\beta}}^{b}
+
 \overbar{\sigma}^{a \dot{\alpha} \beta} \sigma_{\beta \dot{\alpha}}^{b}
\right)  e_a^{~\mu} e_b^{~\nu}  \partial_\mu A \,  \partial_\nu A \, F^*
\nn \\
&=&
16
\left(
-2\eta^{a b}
-2\eta^{a b}
\right)  e_a^{~\mu} e_b^{~\nu}  \partial_\mu A \,  \partial_\nu A \, F^*
\nn \\
&=&
-64 \, (\partial A )^2  \, F^*
\nn \\
\eea
Putting this back into \eqref{f2} and then inserting in \eqref{f1} yields
\bea
\oD^2 \big( (\Phi +  \Phi^\dagger)^{-3}
\D \Phi  \D \Phi  \oD^2 \Phi^\dagger \big)  \big| 
&=&-64 \frac{1}{(A +  A^*)^{3}} \, (\partial A )^2  \, F^* \ .
\nn \\
\eea
The remaining three terms in \eqref{m1} can be evaluated using similar methods. Putting these four component field terms together, and eliminating the $b_{\mu}$ and $M$ auxiliary fields of supergravity, yields the $\mathcal{\bar{L}}_{3, \mathrm{I}}$ contribution to \eqref{A3}.
%%%%%%%%%%%%%%%

\section*{Appendix B: Useful Supergravity Identities}

%%%%%%%%%%%%%%%%%
Here we present a non-exhaustive list of identities necessary for the computations described in Appendix  A and used throughout the paper. 

\noindent The purely superfield results of interest are 
\bea
&& \D_\alpha\D_\beta \D_\gamma \Phi = 
\tfrac{1}{3} ( \{ \D_\alpha, \D_\beta \} \D_\gamma - \{ \D_\alpha, \D_\gamma \} \D_\beta ) \Phi \\
\label{m2}
\nn \\
&& \D_{\alpha} \oD_{\dot{\beta}} \oD_{\dot{\gamma}} \Phi^\dagger =
\oD_{\dot{\eps}} \Phi^\dagger R_{\alpha \dot{\beta}~\dot{\gamma}}^{~~~\dot{\eps}} - 2 i \sigma_{\alpha \dot{\beta}}^e \D_e \oD_{\dot{\gamma}}\Phi^\dagger
- 2i \sigma_{\alpha \dot{\gamma}}^e 
\left( 
T_{\dot{\beta} e}^{~~a} \D_a \Phi^\dagger + T_{\dot{\beta} e}^{~~\eps} \D_{\eps}\Phi^\dagger + \oD^{\dot{\eps}} \Phi^\dagger T_{\dot{\beta} e}^{~~\dot{\eps}}
\right)
\label{m3}
\nn \\
\\
&&\D_{\alpha} \D_{\beta} \oD_{\dot{\alpha}} \D_{\phi} \Phi
=\D_{\alpha} \D_{\beta} \{ \oD_{\dot{\alpha}} , \D_{\phi} \}\Phi
\nn \\
\\
&&\D_{\alpha} \oD_{\dot{\alpha}} \oD_{\dot{\beta}} \D_{\phi} \Phi
=2i \sigma_{\phi \dot{\beta}}^a
\D_{\alpha}
\big(
T_{\dot{\alpha} a}^{~~\eps} \D_{\eps} \Phi 
\big)
\nn \\
\\
&&\D_{\alpha} \oD_{\dot{\alpha}} \oD_{\dot{\gamma} } \oD_{\dot{\delta}} \Phi^\dagger
=
\{ \D_{\alpha} , \oD_{\dot{\alpha}} \} \oD_{\dot{\gamma}} \oD_{\dot{\delta}} \Phi^\dagger
-
\oD_{\dot{\alpha}} \{ \D_{\alpha} , \oD_{\dot{\gamma}}  \} \oD_{\dot{\delta}} \Phi^\dagger
+
\oD_{\dot{\alpha}} \oD_{\dot{\gamma}} \{ \D_{\alpha} , \oD_{\dot{\delta}} \} \Phi^\dagger \ .
\eea
When calculating the lowest component of a superfield expression--indicated by `` \big| ''--we drop all fermions and present the purely bosonic result. The lowest component expressions for the relevant superfields are given by
\bea
&& \D_\alpha \Phi  \big| =0  \ ,    \oD_{\dot{\alpha}} \Phi^\dagger \big| = 0\\
\nn \\
&& \D_\alpha \D_\beta \Phi \big| = - 2 \epsilon_{\alpha \beta} F, \quad \D^2 \Phi \big|= -4 F \\
\nn \\
&& \overbar{\D}_{\dot{\alpha }} \overbar{\D}_{\dot{\beta}} \Phi^\dagger \big| = 2 \epsilon_{\dot{\alpha}\dot{\beta}} F^*,  \quad \overbar{\D}^2 \Phi^\dagger \big| = -4 F^*\\
\nn \\
&& \overbar{\D}_{\dot{\alpha}} \D_\alpha \Phi \big| = - T_{\alpha \dot{\alpha} }^{~~~a} \D_a \Phi\big| = -2i \sigma_{\alpha \dot{\alpha}}^{a} e_a^{~\mu} \partial_\mu A\\
\nn \\
&&  \D_\alpha \overbar{\D}_{\dot{\alpha}} \Phi ^\dagger\big| = -2i \sigma_{\alpha \dot{\alpha}}^{a} e_a^{~\mu} \partial_\mu A^*\\
\nn \\
&& \D_{\alpha} \oD_{\dot{\alpha}} \oD_{\dot{\beta}} \Phi^\dagger \big|= 0
\nn \\
\\
&& \D^2 \oD^2 \Phi^\dagger \big|
=
16 e_a^{~\mu} \D_\mu \hat{D}^a A^* + \tfrac{32}{3} i b^a \hat{D}^a A^* + \tfrac{32}{3} M^*F^*
\\
\nn \\
&&
\oD^2 \oD^2 \Phi^\dagger \big| 
=
\eps^{\dot{\alpha} \dot{\beta}} \eps^{\dot{\gamma} \dot{\delta}}
\oD_{\dot{\alpha}} \oD_{\dot{\beta}} \oD_{\dot{\gamma}} \oD_{\dot{\delta}}\Phi^\dagger \big| 
=
\tfrac{16}{3} F^* M
\eea
\bea
\D_{\alpha} \D_{\beta} \oD_{\dot{\alpha}} \D_{\phi} \Phi \big| 
&=&
-2 i \sigma_{\phi \dot{\alpha}}^a
\big(
- T_{\beta a}^{~~\eps} \big| \D_{\alpha}  \D_{\eps} \Phi \big|
+
 \D_a \D_{\alpha} \D_{\beta} \Phi \big|
\big)
\nn \\
&=&
8 i \sigma_{\phi \dot{\alpha}}^a e_{a}^{~\mu} \partial_\mu F - \tfrac{2}{3} F \sigma_{\phi \dot{\alpha}}^a b_a
\\
\nn \\
\D_{\alpha} \oD_{\dot{\alpha}} \oD_{\dot{\beta}} \D_{\phi} \Phi \big|
&=&
\tfrac{2}{3} M F \sigma_{\phi \dot{\beta}}^a \sigma_{a\alpha \dot{\alpha}}
\eea

Additionally, we find that 
\bea
\D^2 \oD^2 \big( \D \Phi  \D \Phi  \oD^2 \Phi^\dagger \big)  \big|  
&=& 
-2 
\oD_{\dot{\beta}} \D \Phi \big|~ \oD^{\dot{\beta}} \D \Phi \big|~ \D^2 \oD^2 \Phi^\dagger \big|
-4 
\oD_{\dot{\beta}} \D \Phi \big|~ \D^{\alpha} \D \Phi \big|~ \D_{\alpha} \oD^{\dot{\beta}} \Phi^\dagger \big| \nn \\
&+& 4
\D^{\beta} \D \Phi \big|~ \oD_{\dot{\alpha}} \D \Phi \big|~ \D_{\beta} \oD^{\dot{\alpha}}  \oD^2 \Phi^\dagger \big|
-2
\D^{\beta} \D \Phi\big|~ \D_{\beta} \D \Phi \big|~ \oD^2 \oD^2 \Phi^\dagger \big|
\nn \\
&-& 2
\D^2 \oD_{\dot{\alpha}} \D \Phi \big|~ \oD^{\dot{\alpha}} \D \Phi \big|~ \oD^2 \Phi^\dagger \big|
-2
\oD_{\dot{\alpha}} \D \Phi \big|~ \D^2 \oD^{\dot{\alpha}} \D \Phi \big|~ \oD^2 \Phi^\dagger \big| 
\nn \\
\eea
and
\bea
\D^2 \oD^2 \big( \D \Phi  \D \Phi  \oD \Phi^\dagger  \oD \Phi^\dagger)  \big|  
&=&
4 \oD_{\dot{\beta}} \D \Phi \big|~ \oD^{\dot{\beta}} \D \Phi \big|~ \D^{\alpha} \oD \Phi^\dagger \big|~ \D_{\alpha} \oD \Phi^\dagger\big|
-
4 \oD_{\dot{\beta}} \D \Phi \big|~\D^{\alpha}  \D \Phi \big|~\oD^{\dot{\beta}}  \oD \Phi^\dagger \big|~ \D_{\alpha} \oD \Phi^\dagger\big|
\nn \\
&+&
4 \D^{\alpha}\D \Phi \big|~ \oD_{\dot{\beta}} \D \Phi \big|~ \oD^{\dot{\beta}} \oD \Phi^\dagger \big|~ \D_{\alpha}\oD \Phi^\dagger\big|
-
4\D^{\alpha} \D \Phi \big|~ \oD_{\dot{\beta}} \D \Phi \big|~\D_{\alpha} \oD \Phi^\dagger \big|~ \oD^{\dot{\beta}}\oD \Phi^\dagger\big|
\nn \\
&+&
4  \D^{\alpha}\D \Phi \big|~ \D_{\alpha} \D \Phi \big|~ \oD_{\dot{\beta}}\oD \Phi^\dagger \big|~ \oD^{\dot{\beta}}\oD \Phi^\dagger\big| \ .
\nn \\
\eea

\end{document}